%% file: jrnl_preprint_ver.tex
\documentclass[lettersize,journal]{IEEEtran}

\usepackage{amsmath,amssymb,amsfonts}
\usepackage{amsthm}
\usepackage{optidef}

\usepackage{graphicx}
\usepackage[caption=false,font=footnotesize]{subfig}
\usepackage{booktabs}
\usepackage{array}
\usepackage{multirow}
\usepackage{tabularx}
\usepackage{siunitx}
\usepackage{hhline}

\usepackage{algorithmic}

\usepackage{textcomp}
\usepackage{gensymb}
\usepackage{pifont}
\usepackage{url}
\usepackage{stfloats}
\usepackage[T1]{fontenc}

\usepackage{tikz}
\usetikzlibrary{arrows.meta,positioning,calc}

\usepackage[backend=biber,style=ieee]{biblatex}
\usepackage{xr}
\usepackage{hyperref}
\begin{document}
\title{\LARGE \bf
Beyond Reconstruction: A Unified Framework for Trustworthy and Explainable EEG Artifact Removal Evaluation
}
\author{
        Phat~Lam$^{1,*}$,
        Hung~Nguyen$^1$,
        Lam~Pham$^{2,*}$,
        Huy~Phan$^{3}$,
        Dat~Tran$^{4}$,\\
        Ian~McLoughlin$^{5}$,
        Roozbeh Razavi-Far$^{6}$,
        Hung Cao$^{1,*}$
\thanks{$^1$ Analytics Everywhere Lab, University of New Brunswick, Canada.}%
\thanks{$^2$ Austrian Institute of Technology, Austria.}%
\thanks{$^3$ German Research Center for Artificial Intelligence (DFKI), Germany.}%
\thanks{$^4$ FPT University, Vietnam.}%
\thanks{$^5$ Singapore Institute of Technology, Singapore.}%
\thanks{$^6$ Trustworthy and Secure AI Lab, University of New Brunswick, Canada.}%
\thanks{(*) Corresponding Author.}
}
\markboth{Journal of \LaTeX\ Class Files,~Vol.~14, No.~8, November~2026}%
{Shell \MakeLowercase{\textit{et al.}}: A Sample Article Using IEEEtran.cls for IEEE Journals}


\maketitle
\begin{abstract}

Deep-learning methods have achieved strong performance in EEG artifact removal, but their evaluation remains dominated by reconstruction metrics such as error, correlation, and signal-to-noise ratio. These metrics measure whether denoised output resembles clean EEG, but do not verify whether the model removes the intended artifact, preserves neural activity, or provides reliable explanations. 
To address these limitations, we propose TriX-EEG, a three-level evaluation framework for trustworthy and explainable EEG artifact removal: Level 1, Signal Validity (SV), evaluates reconstruction quality and neural band-power preservation; Level 2, Removal Faithfulness (RF), assesses the removed component by quantifying artifact error and  residual artifact/neural leakage; and Level 3, Explanation Reliability (ER), examines model behavior through temporal alignment, causal relevance, and stability. 
To evaluate our proposed framework, we leverage nine published neural denoisers with diverse architectures and three public EEG benchmark datasets derived from EEGDenoiseNet and the PhysioNet Motion Artifact Contaminated EEG Dataset, covering EOG, EMG, and motion artifacts. 
Our experimental results indicate that reconstruction alone is insufficient to characterize trustworthy EEG artifact removal: RF complements it by revealing whether errors stem from residual artifact or suppressed neural activity, whereas ER diverges from both by showing complementary aspects of model explainability, supporting the joint assessment of all three aspects within a unified framework rather than reliance on any single factor. 
Furthermore, our external validation on an independent EEG/EOG dataset and the downstream BCI Competition IV-2a dataset for motor-imagery task shows that TriX-EEG rankings remain broadly consistent under dataset shift, with higher-ranked denoisers generally preserving more task-relevant information or causing less performance degradation under contamination. 
Given these results, TriX-EEG provides a more comprehensive basis for comparing EEG denoisers and encourages models that are not only accurate, but also faithful and explainable.

\indent \textit{Keywords}--- EEG artifact removal, Electrooculography (EOG), Electromyography (EMG), Motion Artifact, Explainable AI, Reconstruction, Evaluation Benchmark.
\end{abstract}
\section{Introduction}
Electroencephalography (EEG) provides a non-invasive measurement of neural activity with high temporal resolution, but it is highly susceptible to artifacts from eye movements (electrooculography, EOG), muscle activity (electromyography, EMG), and motion. These artifacts can distort neural rhythms and reduce the reliability of downstream analysis~\cite{review_01, review_03}. Removing these artifacts is challenging. Indeed, insufficient suppression leaves contamination in the signal, whereas excessive correction
may attenuate neural activity relevant to downstream analysis~\cite{eeg_left_along}.
These challenges highlight the need to evaluate artifact removal not only by its
apparent effectiveness, but also by the information it
preserves.

Deep neural networks have become increasingly popular for EEG artifact removal because they can learn nonlinear mappings from contaminated EEG to clean EEG and often achieve strong reconstruction performance~\cite{eeg_denoising_survey}.
The introduction of EEGDenoiseNet \cite{eegdenoisenet} has facilitated systematic
comparisons under controlled EOG and EMG contamination, with performance commonly quantified using temporal and spectral relative root-mean-square errors (RRMSE), correlation coefficient (CC), and signal-to-noise ratio (SNR) improvement.
These metrics provide essential measures of reconstruction quality, but primarily summarize the overall agreement between denoised and reference signals. Consequently, they do not explicitly describe the structure of the remaining
error, including its relationship to residual artifact, unintended neural attenuation, or frequency-specific distortion. Moreover, a favorable reconstruction may still suppress neural activity or retain part of the artifact, and does not necessarily establish that the denoised signal retains
the information required by downstream EEG applications
\cite{eeg_left_along}, and therefore may not represent trustworthy artifact removal.

In short, standard reconstruction metrics primarily quantify the discrepancy between the denoised output and the clean EEG reference~\cite{eegdenoisenet}, providing an overall measure of reconstruction fidelity without explicitly distinguishing residual artifact from unintended neural-signal suppression. A more informative evaluation should therefore examine not only the reconstructed EEG, but also the signal removed
by the denoiser. 
Although reconstruction and removal errors are coupled, evaluating the removed component against the paired artifact and clean EEG references provides a complementary perspective on denoising behavior. 
For instance, denoisers with similar reconstruction errors may differ in their relative artifact suppression and neural-signal preservation. 
Thus, examining the removed component against paired neural and artifact references can provide additional diagnostic information about the nature of denoising errors, complementing conventional
output-based metrics. 
This distinction is particularly relevant when neural activity and artifacts overlap in frequency or when subtle signal distortions may affect subsequent analysis \cite{wiener_filter_artifact,
is_brain_reserved,suppresing_motion_artifacts}.

Beyond signal quality, the interpretability of learned denoising operations raises a further evaluation challenge. Explainable artificial intelligence (XAI) methods are increasingly applied to EEG models to identify influential temporal, spectral, or spatial features
\cite{XAI_eeg_survey,XAI_eeg_survey_02}. However, visually plausible attribution maps do not necessarily provide faithful explanations of model behavior. Some saliency methods can be insensitive to model parameters or data relationships, while attribution quality may vary substantially across time-series models
\cite{saliency_map_sanity, saliency_time_series}. Existing quantitative XAI evaluations have primarily focused on classification tasks \cite{XAI_benchmark_time_series}, where explanations are typically associated with a specific class prediction. In contrast, EEG denoising maps a contaminated waveform to a reconstructed waveform, so the model operation being explained should be defined explicitly. Even with a well-defined target, highlighting influential input regions does not establish that a denoiser attends to the artifact rather than to neural activity, nor that the attribution actually drives its behavior. A reliable explanation of a denoiser should therefore satisfy complementary properties. It should be temporally aligned with the true artifact. The model’s spectral sensitivity should be consistent with the artifact spectrum. Also, removing the highlighted inputs should causally change the denoising operation. Finally, the attribution should remain stable under small, non-semantic input perturbations. 

Motivated by these limitations, we propose \textit{TriX-EEG}, a three-level evaluation framework that systematically assesses EEG artifact removal from complementary perspectives. \textbf{Level 1 (Signal Validity, SV)} evaluates how accurately clean EEG is reconstructed, considering temporal and spectral reconstruction quality and neural band-power preservation. \textbf{Level 2 (Removal Faithfulness, RF)} characterizes the removed component and its errors relative to neural and artifact references through artifact similarity, energy agreement, reconstruction error, and conditional estimates of artifact gain and neural leakage. \textbf{Level 3 (Explanation Reliability, ER)} assesses whether explanations of the model operation exhibit temporal alignment with artifact activity, spectral sensitivity consistent with the artifact spectrum, intervention-based relevance, and stability under small input perturbations. The three levels are retained as diagnostic profiles and integrated into an overall TriX-EEG Denoising Score (TDS) using a geometric mean. Together, they extend conventional reconstruction-based evaluation by examining not only the quality of the denoised signal, but also the characteristics of the removed component and the reliability of explanations of the denoising operation.

The main contributions of this work are as follows:
\begin{itemize}

    \item \textbf{A structured multi-level evaluation framework.}
    We introduce \textit{TriX-EEG}, which integrates
    conventional reconstruction measures, removed-component
    diagnostics, and quantitative attribution metrics for EEG artifact removal.
    The framework provides interpretable level-wise profiles
    alongside a unified summary score for systematic model
    comparison.
    
    \item  \textbf{Comprehensive benchmarking and external
    validation.} We evaluate nine neural denoisers across EOG, EMG,
    and motion-artifact benchmarks, followed by cross-domain
    validation on an independent EEG/EOG dataset and
    downstream motor-imagery classification on BCI
    Competition IV-2a dataset. These experiments examine the
    agreement between evaluation dimensions, the
    transferability of model rankings, and their relationship
    to downstream task-level performance under controlled contamination.

    \item \textbf{Systematic metric validation through controlled sanity checks.}
    We investigate the behavior and limitations of the proposed evaluation metrics under controlled failure conditions, including model randomization, incorrect artifact references, shuffled attributions, and clean-input evaluation. We further examine the sensitivity of the framework to evaluation settings, assessing the robustness and interpretability of its component metrics and aggregate scores.

    \item \textbf{Reproducible evaluation resources.}
    We release an open-source implementation of the
    proposed evaluation framework, component metrics,
    and benchmarking procedures at
    \url{https://github.com/Analytics-Everywhere-Lab/TriX-EEG}.

\end{itemize}

Our experiments reveal the complementarity between SV and RF, while their component profiles provide different diagnostic perspectives on denoising errors. In contrast, ER differs markedly from signal-based evaluations, suggesting possible different evaluation aspects and emphasizing the importance of interpreting attribution measures alongside denoising quality. External validation further demonstrates broad cross-domain ranking consistency and associations between TDS and relative downstream performance. Together, these findings motivate evaluation beyond reconstruction scores toward a more
explicit characterization of what denoisers preserve, what they remove, and how their operations can be examined.

\section{Related Work}
\label{sec:related_work}

\subsection{EEG Artifact Removal Methods}
EEG artifact removal has traditionally relied on signal processing and source-separation techniques. Regression-based methods estimate artifact contributions from reference
recordings, such as EOG channels, while filtering and
time--frequency decomposition methods exploit differences in the spectral or temporal characteristics of neural and artifactual activity~\cite{review_removal_old_1,review_removal_old_2}. Blind source separation, particularly independent component analysis (ICA), has been widely used to identify and remove components associated with ocular, muscular, and other non-neural sources~\cite{jung2000removing}. Automated component classification approaches, including MARA and ICLabel, have further reduced the manual effort required for artifact identification~\cite{mara,iclabel}. Despite their practical utility, conventional approaches
may rely on assumptions about source independence,
reference-signal availability, or artifact characteristics that are difficult to satisfy consistently across recording conditions. 

Deep learning methods have increasingly been investigated for EEG artifact removal because of their ability to learn nonlinear mappings between contaminated and reference EEG signals \cite{eeg_denoising_survey,wu2026_new_survey}. EEGDenoiseNet \cite{eegdenoisenet} established a benchmark for evaluating learned denoisers under controlled EOG and EMG contamination. Subsequent studies have explored diverse approaches to this task, e.g., convolutional/recurrent architectures, encoder--decoder networks, generative adversarial models, Transformer-based models and specialized denoising designs~\cite{wu2026_new_survey, single_channel_survey}. While architectural improvements have advanced artifact suppression in benchmark settings, evaluating a denoiser requires more than identifying the model with the lowest reconstruction error. In particular, the diversity of denoising architectures raises questions about whether comparable output quality reflects similar removal behavior and whether learned operations preserve information relevant to downstream EEG analysis. These questions motivate evaluation
protocols that extend beyond conventional reconstruction benchmarks.

\subsection{Evaluation of EEG Denoising Quality}
EEG artifact-removal methods are commonly evaluated using reconstruction metrics, including temporal and
spectral RRMSE, CC, and SNR improvement \cite{eegdenoisenet}. EEGDenoiseNet \cite{eegdenoisenet}
provides a widely used semi-simulated benchmark in which clean EEG segments are mixed with recorded EOG or EMG artifacts at controlled SNR levels, enabling reproducible comparisons with known signal components. 

Beyond waveform reconstruction, preserving neural activity is critical for the reliability of artifact removal. 
Prior studies have also examined the trade-off between artifact suppression and neural preservation. For example,  \cite{chang2019evaluation_asr} evaluated the removal of
artifact- and brain-associated components under different artifact subspace reconstruction settings, while~\cite{eeg_left_along} showed that EEG preprocessing can reduce the sensitivity of event-related potential analyses. These findings highlight that favorable signal-cleaning performance does not necessarily imply preservation of task-relevant information, motivating evaluation through
both signal-level measures and downstream tasks.

Another important consideration is the characterization of the signal removed by a denoiser. Aggregate reconstruction metrics do not explicitly distinguish how this error relates
to the artifact and clean EEG references. Analyzing the removed component can therefore provide complementary diagnostic information about residual artifact, artifact-gain errors, and clean-signal-aligned attenuation. Building on prior work in neural-preservation assessment, TriX-EEG
incorporates reference-guided removal diagnostics alongside conventional reconstruction measures and examines their relationship with downstream EEG decoding performance.

\subsection{Quantitative Evaluation of Model Explanations}

XAI methods have been widely investigated for identifying input features that influence model predictions in EEG analysis~\cite{XAI_eeg_survey_02}. However, visually
plausible attribution maps do not necessarily provide faithful explanations. Certain saliency methods can remain insensitive to model parameter or data randomization, highlighting the need for quantitative evaluation~\cite{saliency_map_sanity}. Perturbation-based methods assess attribution relevance by measuring changes in model outputs when influential features are modified or removed
\cite{pertubation_XAI_1,pertubation_XAI_2}, while
complementary approaches examine explanation robustness, sensitivity, and infidelity
\cite{robustness_XAI, fidelity_XAI}.

Evaluating explanations for time-series models introduces additional challenges because relevant information may vary jointly across temporal and feature dimensions. For example, \cite{saliency_time_series} demonstrated limitations of conventional saliency methods in identifying important temporal features, while
\cite{XAI_benchmark_time_series} systematically assessed
post-hoc explanations for time-series classification. Although these studies establish important evaluation
principles, representative benchmarks have primarily
focused on classification. 
In contrast, EEG denoising is a signal-to-signal
regression task, for which the operation being explained must be explicitly defined rather than inferred from a target-class prediction.

For EEG denoising, explaining the reconstructed waveform or the removed signal does not inherently establish whether the underlying operation correctly suppresses artifacts while preserving neural activity. Moreover, attribution
alignment, intervention-based relevance, and stability capture distinct properties of explanation behavior rather than providing definitive evidence of correct artifact separation. Therefore, TriX-EEG defines an explicit removed-signal energy target and incorporates complementary
temporal, spectral, intervention-based, and stability
assessments. These explanation diagnostics are considered alongside reconstruction and removal quality to provide a more comprehensive characterization of denoising behavior.

\begin{figure*}[t]
    \centering
    \includegraphics{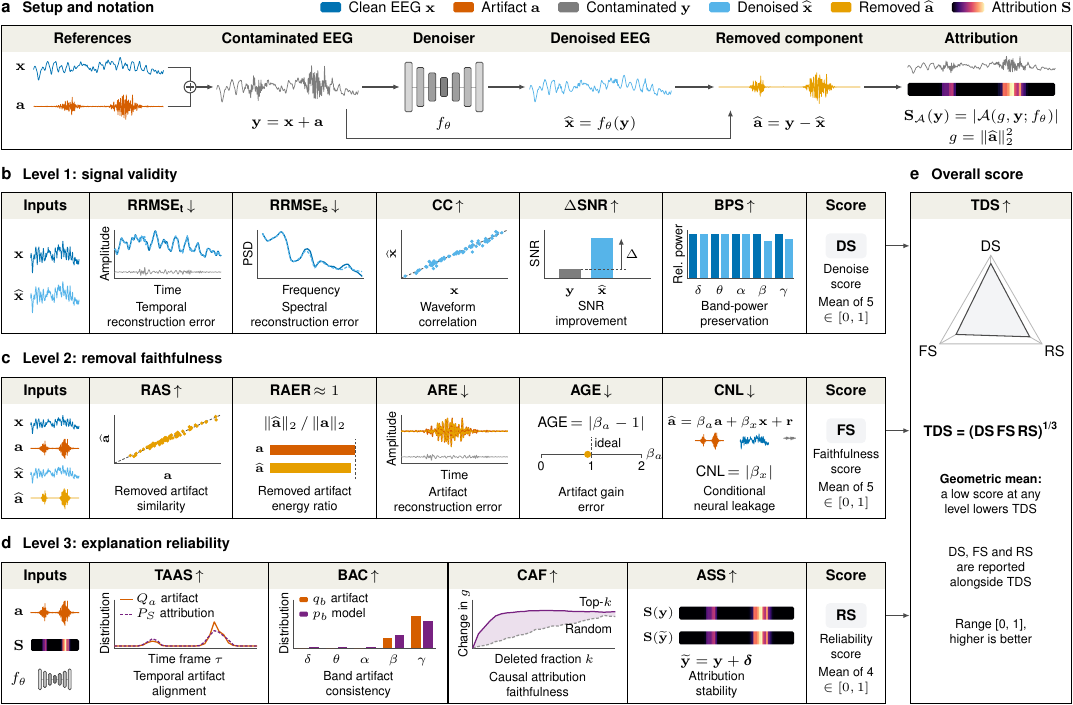}
    \caption{\textbf{Overview of the TriX-EEG evaluation framework.} (a)~A denoiser
    $f_\theta$ maps the contaminated EEG $\mathbf{y}=\mathbf{x}+\mathbf{a}$ to
    $\widehat{\mathbf{x}}$; the removed component $\widehat{\mathbf{a}}$ and
    the attribution $\mathbf{S}(\mathbf{y})$ of the removed energy $g$ are
    evaluated against the references $\mathbf{x}$ and $\mathbf{a}$.
    (b)~Level~1 compares $\widehat{\mathbf{x}}$ with $\mathbf{x}$.
    (c)~Level~2 compares $\widehat{\mathbf{a}}$ with $\mathbf{a}$.
    (d)~Level~3 tests whether the explanation follows the artifact in time
    (TAAS) and frequency (BAC), is causally relevant (CAF) and is stable
    under a small perturbation $\boldsymbol{\delta}$ (ASS). (e)~The level
    scores are combined into TDS by a geometric mean. Arrows give the
    preferred direction of each metric; solid lines are references, dashed
    lines model estimates. Mini-plots are schematic and computed on simulated
    signals.}
    \label{fig:trix-eeg-overall}
    \vspace{-0.3cm}
\end{figure*}

\section{TriX-EEG: Proposed Framework}
\label{sec:trix_eeg}

\subsection{Problem Definition and Notation}
\label{subsec:problem_definition}

Let $\mathbf{y}\in\mathbb{R}^{C\times T}$ denote a contaminated EEG signal
with $C$ channels and $T$ time samples. We assume that it can be represented as
the sum of a clean neural component $\mathbf{x}\in\mathbb{R}^{C\times T}$ and
an artifact component $\mathbf{a}\in\mathbb{R}^{C\times T}$:
\begin{equation}
    \mathbf{y}=\mathbf{x}+\mathbf{a}.
    \label{eq:additive_observation}
\end{equation}

Given $\mathbf{y}$, a denoising model $f_\theta$ estimates the clean EEG as
\begin{equation}
    \widehat{\mathbf{x}}=f_\theta(\mathbf{y}),
    \label{eq:trix_prediction}
\end{equation}
where $\theta$ denotes the learned model parameters. The signal removed by the
model is consequently defined as
\begin{equation}
    \widehat{\mathbf{a}}
    =\mathbf{y}-\widehat{\mathbf{x}}.
    \label{eq:trix_removed_component}
\end{equation}
Ideally, $\widehat{\mathbf{x}}$ should approximate $\mathbf{x}$ while
$\widehat{\mathbf{a}}$ should approximate $\mathbf{a}$. These objectives are
related but not equivalent: a denoised output may resemble the clean reference
even when the model removes an incorrect mixture of artifact and neural
activity.

\textit{TriX-EEG} is a three-level evaluation framework designed to evaluate complementary aspects
of EEG denoising: SV, RF, and
ER. 
The evaluation framework comprises 14 component metrics distributed across the three
levels. Each component is expressed as, or transformed into, a bounded utility
in $[0,1]$, where a larger value consistently indicates better behavior.
Level~1 (SV) evaluates the agreement between the denoised
output $\widehat{\mathbf{x}}$ and the clean reference $\mathbf{x}$ in both the
temporal and spectral domains. Level~2 (RF) examines
whether the removed component $\widehat{\mathbf{a}}$ corresponds to the true
artifact $\mathbf{a}$ while also accounting for neural leakage and residual artifact
contamination. Level~3 (ER) evaluates the temporal alignment, spectral sensitivity, intervention-based relevance, and stability of explanations of the denoising operation. The level-wise scores are reported alongside the aggregate score to preserve diagnostic information.

For Level~3, we define a scalar target representing the total energy removed by the denoiser:
\begin{equation}
g(\mathbf{y})
=
\left||
\mathbf{y}-f_\theta(\mathbf{y})
|\right|_2^2
=
\left||\widehat{\mathbf{a}}|\right|_2^2.
\label{qg:removed_energy_target}
\end{equation}
Given an input-attribution method $\mathcal{A}$, the corresponding nonnegative attribution map is defined as
\begin{equation}
\mathbf{S}_{\mathcal{A}}(\mathbf{y})
=
\left|
\mathcal{A}(g,\mathbf{y};f_\theta)
\right|,
\label{eq:attribution_map}
\end{equation}
where $\mathcal{A}$ assigns an importance value to each input sample with respect to the selected scalar target. The resulting map characterizes the input regions associated with the model's removed-signal energy.

All component metrics are first calculated independently for each EEG window and then macro-averaged over the evaluation set. 

\subsection{Level 1: Signal Validity}
\label{subsec:level1}

Level~1 asks whether the denoised output remains a valid estimate of clean EEG. 
It combines standard reconstruction metrics: temporal and spectral reconstruction, correlation,
noise suppression, and preservation of physiologically meaningful rhythms~\cite{eegdenoisenet}.
\textbf{The temporal and spectral RRMSE} are defined as:
\begin{align}
    \mathrm{RRMSE}_{t}
    &=\frac{\|\widehat{\mathbf{x}}-\mathbf{x}\|_2}
            {\|\mathbf{x}\|_2+\epsilon},
    \label{eq:rrmse_t}\\
    \mathrm{RRMSE}_{s}
    &=\frac{\|\mathrm{PSD}(\widehat{\mathbf{x}})-
                    \mathrm{PSD}(\mathbf{x})\|_2}
            {\|\mathrm{PSD}(\mathbf{x})\|_2+\epsilon},
    \label{eq:rrmse_s}
\end{align}
where $\mathrm{PSD}(\cdot)$ is the one-sided FFT periodogram. 

We transform them into utility metrics as:

\begin{equation}
    U_t=1-\operatorname{clip}(\mathrm{RRMSE}_t,0,1)
\end{equation}
\begin{equation}
    U_s=1-\operatorname{clip}(\mathrm{RRMSE}_s,0,1)
\end{equation}

Waveform agreement is measured by the \textbf{Pearson correlation}:
\begin{equation}
    \mathrm{CC}=\operatorname{corr}
    (\widehat{\mathbf{x}},\mathbf{x}),
    \qquad U_{\mathrm{CC}}=\operatorname{clip}(\mathrm{CC},0,1).
    \label{eq:level1_cc}
\end{equation}
To quantify denoising gain relative to the input corruption, we use \textbf{SNR improvement}:

\begin{equation}
    \Delta\mathrm{SNR}
    =\mathrm{SNR}_{\mathrm{out}}-\mathrm{SNR}_{\mathrm{in}},
\end{equation}
\begin{equation}
    U_{\Delta\mathrm{SNR}}=
    \operatorname{clip}\!\left(\frac{\Delta\mathrm{SNR}}{\Delta SNR_{\text{max}}},0,1\right).
    \label{eq:delta_snr}
\end{equation}
Regarding spectral error, we leverage band-specific relative power error over $\mathcal{B}$ frequency bands:
\begin{equation}
    E_b^{\mathrm{BP}}=
    \frac{|P_b(\widehat{\mathbf{x}})-P_b(\mathbf{x})|}
         {P_b(\mathbf{x})+\epsilon},
    \qquad
    \overline{E}^{\mathrm{BP}}=
    \frac{1}{|\mathcal{B}|}\sum_{b\in\mathcal{B}}E_b^{\mathrm{BP}}.
    \label{eq:band_power_error}
\end{equation}
\textbf{The Band-Power Preservation Score (BPS)} is computed as
\begin{equation}
    \mathrm{BPS}=1-
    \operatorname{clip}\!\left(\frac{\overline{E}^{\mathrm{BP}}}
    {E_{\max}^{\mathrm{BP}}},0,1\right).
    \label{eq:bps}
\end{equation}
Finally, the Level-1 Denoise Score (DS) is the transparent arithmetic summary
\begin{equation}
    \mathrm{DS}=\frac{1}{5}
    \left(U_t+U_s+U_{\mathrm{CC}}+U_{\Delta\mathrm{SNR}}+
    \mathrm{BPS}\right).
    \label{eq:ds}
\end{equation}
The component profile is inspected together with DS because equal means
can arise from different temporal, spectral, and band-preservation trade-offs.

\subsection{Level 2: Removal Faithfulness}
\label{subsec:level2}

Good reconstruction alone cannot determine whether a model removed the true
artifact or inadvertently suppressed neural activity~\cite{eeg_denoising_survey}. Level~2 therefore
evaluates the removed component in Eq.~\eqref{eq:trix_removed_component}.

First, \textbf{Removed Artifact Similarity (RAS)} measures morphological agreement via correlation operation as 
\begin{equation}
    \mathrm{RAS}=\max\!\left(0,
    \operatorname{corr}(\widehat{\mathbf{a}},\mathbf{a})\right),
    \label{eq:ras}
\end{equation}
while the \textbf{Removed Artifact Energy Ratio (RAER)} evaluates whether an appropriate
amount of energy was removed: \begin{equation}
    \mathrm{RAER}=
    \frac{\|\widehat{\mathbf{a}}\|_2}{\|\mathbf{a}\|_2+\epsilon},
    \qquad
    \label{eq:raer}
\end{equation}
  \begin{equation}
    U_{\mathrm{RAER}}
    =
    1-
    \operatorname{clip}\left(
        \frac{|\mathrm{RAER}-1|}
             {E_{\max}^{\mathrm{RAER}}},
        0,1
    \right).
    \label{eq:raer}
\end{equation}
RAS and RAER can still reward a removed signal with approximately correct
correlation or energy but substantial pointwise mismatch. We consequently add
\textbf{Artifact Reconstruction Error (ARE)},
\begin{equation}
    \mathrm{ARE}=
    \frac{\|\widehat{\mathbf{a}}-\mathbf{a}\|_2}
         {\|\mathbf{a}\|_2+\epsilon},
    \qquad
    U_{\mathrm{ARE}}=1-
    \operatorname{clip}(\mathrm{ARE},0,1),
    \label{eq:are}
\end{equation}
Together, the first three metrics penalize morphological, amplitude, and alignment errors between true artifact and removed component.

However, RAS and RAER alone cannot reliably distinguish faithful artifact suppression from unintended neural signal removal. Given the paired clean EEG $\mathbf{x}$ and reference artifact $\mathbf{a}$, we analyze both aspects by decomposing the removed component as
\begin{equation}
    \widehat{\mathbf{a}}
    =
    \beta_a\mathbf{a}
    +
    \beta_x\mathbf{x}
    +
    \mathbf{r},
    \label{eq:joint_removal_decomposition}
\end{equation}
where $\beta_a$ represents the conditional artifact gain, $\beta_x$ quantifies neural signal leakage, and $\mathbf{r}$ denotes the component unexplained by either reference. Under ideal artifact removal, $\widehat{\mathbf{a}}=\mathbf{a}$, corresponding to $\beta_a=1$, $\beta_x=0$, and $\mathbf{r}=\mathbf{0}$.

To estimate the coefficients, we jointly regress the removed component against the artifact and clean EEG references using Ridge regression~\cite{ridge}:
\begin{equation}
    (\beta_a,\beta_x)
    =
    \arg\min_{b_a,b_x}
    \left\|
        \widehat{\mathbf{a}}
        -b_a\mathbf{a}
        -b_x\mathbf{x}
    \right\|_2^2
    +
    \lambda(b_a^2+b_x^2),
    \label{eq:joint_removal_regression}
\end{equation}
where Ridge regularization improves numerical stability, particularly when the two references are correlated. The regularization strength is defined as
\begin{equation}
    \lambda
    =
    10^{-6}
    \cdot
    \frac{1}{2}
    \operatorname{tr}\left(
        \mathbf{X}^{\top}\mathbf{X}
    \right),
    \qquad
    \mathbf{X}=[\mathbf{a},\mathbf{x}].
\end{equation}
Joint regression allows artifact gain and neural leakage to be assessed conditionally, without requiring the reference signals to be orthogonal. Based on the estimated coefficients, we define two complementary metrics.

\textbf{Artifact Gain Error (AGE): }
AGE quantifies the deviation of the conditional artifact gain from its ideal value:
\begin{equation}
    \mathrm{AGE}
    =
    |\beta_a-1|.
    \label{eq:age_raw}
\end{equation}
A lower AGE indicates that the removed component contains the reference artifact with approximately the correct gain, after accounting for the clean EEG contribution. Its normalized, higher-is-better utility is defined as
\begin{equation}
    U_{\mathrm{AGE}}
    =
    1-
    \operatorname{clip}\left(
        \frac{\mathrm{AGE}}
             {E_{\max}^{\mathrm{AGE}}},
        0,1
    \right).
    \label{eq:age}
\end{equation}

\textbf{Conditional Neural Leakage (CNL).}
CNL measures the magnitude of the clean EEG contribution to the removed component, conditioned on the artifact reference:
\begin{equation}
    \mathrm{CNL}
    =
    |\beta_x|.
    \label{eq:cnl_raw}
\end{equation}
A lower CNL indicates less unintended removal of neural signal that is linearly attributable to the clean EEG reference. Its normalized utility is given by
\begin{equation}
    U_{\mathrm{CNL}}
    =
    1-
    \operatorname{clip}\left(
        \frac{\mathrm{CNL}}
             {E_{\max}^{\mathrm{CNL}}},
        0,1
    \right).
    \label{eq:cnl}
\end{equation}

Together, AGE and CNL conditionally characterize two complementary aspects of RF: whether the intended artifact is removed with the correct gain and whether neural activity is inadvertently removed. Unlike RAS, these metrics explicitly distinguish the contributions of artifact and clean EEG references within the estimated removed component.

Finally, the Level-2 Faithfulness Score (FS) combines five complementary,
higher-is-better criteria using an equal-weight arithmetic mean:
\begin{equation}
    \mathrm{FS}
    =
    \frac{1}{5}
    \left(
        \mathrm{RAS}
        +U_{\mathrm{RAER}}
        +U_{\mathrm{ARE}}
        +U_{\mathrm{AGE}}
        +U_{\mathrm{CNL}}
    \right).
    \label{eq:fs}
\end{equation}
A high FS therefore requires the removed component to reproduce the reference
artifact in morphology, total energy, pointwise structure, and conditional
artifact gain while limiting clean-signal leakage.




\subsection{Level 3: Explanation Reliability}
\label{subsec:level3}

Level~3 evaluates operational properties of explanations
of the denoising operation through four complementary
criteria: Temporal Artifact Alignment Score (TAAS), Band
Artifact Consistency (BAC), Causal Attribution Faithfulness
(CAF), and Attribution Stability Score (ASS).
TAAS, CAF, and ASS use the attribution map
$\mathbf{S}_{\mathcal A}(\mathbf{y})$ defined in
Eq.~\eqref{eq:attribution_map}, whereas
BAC directly evaluates model responses to spectral
interventions without relying on an attribution method.

\textbf{TAAS:}
TAAS measures the temporal correspondence between
attribution magnitude and reference artifact energy.
Let $Z(\tau,f)$ denote the short-time Fourier transform
(STFT) of the reference artifact $\mathbf{a}$, where
$\tau$ and $f$ index time frames and frequencies,
respectively. The normalized temporal artifact-energy
distribution is defined as
\begin{equation}
    Q_a(\tau)=
    \frac{
        \sum_{f\in\mathcal{F}_a}|Z(\tau,f)|^2
    }{
        \sum_u\sum_{f\in\mathcal{F}_a}|Z(u,f)|^2+\epsilon
    },
    \label{eq:artifact_temporal_distribution}
\end{equation}
where $\mathcal{F}_a$ denotes the evaluated artifact
frequency range.

The attribution map
$\mathbf{S}_{\mathcal A}(\mathbf{y})$ is aggregated
over the corresponding STFT frames to obtain
a non-negative temporal attribution magnitude
$M_{\mathcal A}(\tau)$. Its normalized distribution is
\begin{equation}
    P_{\mathcal A}(\tau)=
    \frac{
        M_{\mathcal A}(\tau)
    }{
        \sum_u M_{\mathcal A}(u)+\epsilon
    }.
    \label{eq:attribution_temporal_distribution}
\end{equation}
\begin{figure}[t]
    \centering
    \includegraphics{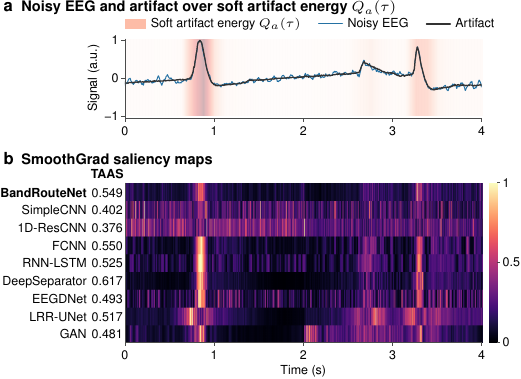}
    \caption{An example of Soft Artifact Energy $Q_a(\tau)$ and Model Attribution alignment $M_{\mathcal A}(\tau)$ on two concatenated windows, corresponding with SmoothGrad saliency and TAAS scores for the compared models on EEGDenoiseNet-EOG dataset.}
    \label{fig:saliency-eog}
    \vspace{-0.3cm}
\end{figure}
\begin{figure}[t]
    \centering
    \includegraphics{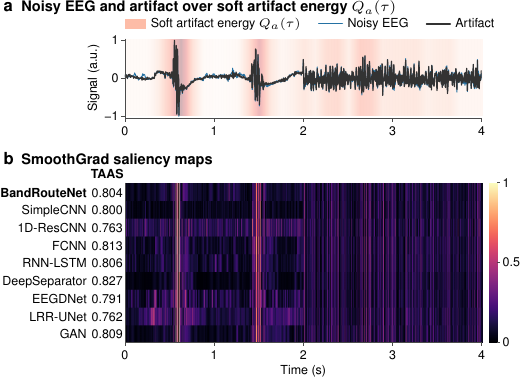}
    \caption{An example of Soft Artifact Energy $Q_a(\tau)$ and Model Attribution alignment $M_{\mathcal A}(\tau)$ on two concatenated windows, corresponding with SmoothGrad saliency and TAAS scores for the compared models on EEGDenoiseNet-EMG dataset.}
    \label{fig:saliency-emg}
    \vspace{-0.3cm}
\end{figure}
TAAS measures the agreement between these temporal
distributions using
\begin{equation}
    \mathrm{TAAS}
    =
    1-\frac{1}{2}
    \sum_{\tau}
    \left|
        P_{\mathcal A}(\tau)-Q_a(\tau)
    \right|.
    \label{eq:taas}
\end{equation}

TAAS provides a soft measure of temporal alignment
without requiring binary artifact masks or explicit
onset--offset annotations. Examples using SmoothGrad
attributions \cite{smoothgrad} for EOG and EMG artifacts
are shown in Figs.~\ref{fig:saliency-eog}
and~\ref{fig:saliency-emg}, respectively.
\begin{table*}[t]
    \centering
    \caption{
        Comparison of Performance of SV Metrics (Level 1) across models
        (Best Results are in Bold; Second-best Ones are Underlined).
    }
    \label{tab:benchmark_results}
    \setlength{\tabcolsep}{4pt}
    \renewcommand{\arraystretch}{1.05}

    \resizebox{0.95\textwidth}{!}{%
    \begin{tabular}{l cccc cccc cccc}
        \hline

        \multirow{2}{*}{\textbf{Method}}
        & \multicolumn{4}{c}{\textbf{EEGDenoiseNet-EOG}}
        & \multicolumn{4}{c}{\textbf{EEGDenoiseNet-EMG}}
        & \multicolumn{4}{c}{\textbf{EEG-Motion}} \\

        \cmidrule(lr){2-5}
        \cmidrule(lr){6-9}
        \cmidrule(lr){10-13}

        &
        \textbf{RRMSE$_{\text{t}}$} $\downarrow$
        & \textbf{RRMSE$_{\text{s}}$} $\downarrow$
        & \textbf{CC} $\uparrow$
        & \textbf{$\Delta$ SNR} $\uparrow$

        & \textbf{RRMSE$_{\text{t}}$} $\downarrow$
        & \textbf{RRMSE$_{\text{s}}$} $\downarrow$
        & \textbf{CC} $\uparrow$
        & \textbf{$\Delta$ SNR} $\uparrow$

        & \textbf{RRMSE$_{\text{t}}$} $\downarrow$
        & \textbf{RRMSE$_{\text{s}}$} $\downarrow$
        & \textbf{CC} $\uparrow$
        & \textbf{$\Delta$ SNR} $\uparrow$ \\
        \hline

        FCNN~\cite{eegdenoisenet}
        & $0.5570$ & $0.5865$ & $0.8111$ & $10.8011$
        & $\underline{0.6176}$ & $0.6687$ & $\mathbf{0.7849}$ & $\underline{9.6891}$
        & $0.8553$ & $1.0593$ & $0.4827$ & $-0.9128$ \\

        Simple-CNN~\cite{eegdenoisenet}
        & $0.4437$ & $0.4432$ & $0.8847$ & $12.6351$
        & $0.7215$ & $0.7263$ & $0.7255$ & $8.4879$
        & $0.7102$ & $1.0616$ & $0.7508$ & $1.1927$ \\

        1D-ResCNN~\cite{1D-ResCNN}
        & $\underline{0.4184}$ & $\underline{0.4046}$ & $0.8966$ & $\underline{13.1761}$
        & $0.7510$ & $0.7093$ & $0.7076$ & $8.2550$
        & $0.7262$ & $1.0736$ & $0.7342$ & $0.9455$ \\

        RNN-LSTM~\cite{eegdenoisenet}
        & $0.6903$ & $0.7231$ & $0.7176$ & $8.6099$
        & $0.6758$ & $0.7215$ & $0.7384$ & $8.8269$
        & $0.7751$ & $1.0000$ & $0.6184$ & $0.0798$ \\

        EEGDNet~\cite{EEGDnet}
        & $0.4406$ & $0.4125$ & $0.8907$ & $12.4712$
        & $0.6554$ & $0.6196$ & $0.7456$ & $9.0505$
        & $0.6641$ & $\underline{0.8578}$ & $0.7485$ & $1.5212$ \\

        Deep Separator~\cite{deepseparator}
        & $0.4944$ & $0.5484$ & $0.8658$ & $11.3920$
        & $0.6982$ & $0.6899$ & $0.7396$ & $8.6998$
        & $0.6249$ & $0.8693$ & $0.7994$ & $2.3499$ \\

        LRR-UNet~\cite{lrr_unet}
        & $0.4322$ & $0.4311$ & $\underline{0.8967}$ & $12.7577$
        & $0.6415$ & $\underline{0.5827}$ & $0.7564$ & $9.3746$
        & $\underline{0.6060}$ & $\mathbf{0.8339}$ & $\underline{0.8123}$ & $\underline{2.5486}$ \\

        GAN~\cite{denoise_gan}
        & $0.6184$ & $0.5675$ & $0.7909$ & $9.5733$
        & $0.8048$ & $0.6870$ & $0.6460$ & $7.3298$
        & $0.6921$ & $1.2168$ & $0.7678$ & $2.0451$ \\

        BandRouteNet~\cite{brn}
        & $\mathbf{0.3831}$ & $\mathbf{0.3797}$ & $\mathbf{0.9156}$ & $\mathbf{13.9819}$
        & $\mathbf{0.5962}$ & $\mathbf{0.5276}$ & $\underline{0.7802}$ & $\mathbf{10.0041}$
        & $\mathbf{0.5949}$ & $0.9141$ & $\mathbf{0.8286}$ & $\mathbf{2.8827}$ \\

        \hline
    \end{tabular}%
    }
    \vspace{-0.2cm}
\end{table*}

\textbf{BAC:}
BAC evaluates whether the model is sensitive to the
same frequency bands that dominate the true artifact. For each frequency band
$b\in\mathcal{B}$, we calculate $\mathbf{y}_{-b}$ by suppressing band $b$ in
the contaminated input $\mathbf{y}$. The relative change in the model output is
then defined as
\begin{equation}
    I_b =
    \frac{
        \left\|f_\theta(\mathbf{y})
        - f_\theta(\mathbf{y}_{-b})\right\|_2
    }{
        \left\|f_\theta(\mathbf{y})\right\|_2+\epsilon
    }.
    \label{eq:bac_band_importance}
\end{equation}
The responses are normalized across all evaluated bands to form the
model-sensitivity distribution
\begin{equation}
    p_b =
    \frac{I_b}
    {\sum_{k\in\mathcal{B}} I_k+\epsilon}.
    \label{eq:bac_model_distribution}
\end{equation}

In parallel, the spectral distribution of the ground-truth artifact is
calculated from its band powers:
\begin{equation}
    q_b =
    \frac{P_b(\mathbf{a})}
    {\sum_{k\in\mathcal{B}}P_k(\mathbf{a})+\epsilon},
    \label{eq:bac_artifact_distribution}
\end{equation}
where $P_b(\mathbf{a})$ denotes the power of artifact $\mathbf{a}$ within band
$b$. BAC measures the similarity between the model-sensitivity distribution
$\{p_b\}$ and the artifact-power distribution $\{q_b\}$ 
\begin{equation}
    \mathrm{BAC}
    =
    1-\frac{1}{2}
    \sum_{b\in\mathcal{B}}
    \left|p_b-q_b\right|.
    \label{eq:bac}
\end{equation}
BAC complements attribution-based assessments by
characterizing the agreement between model sensitivity to frequency-band suppression and the reference artifact spectrum.
\begin{figure}[t]
    \centering
    \includegraphics{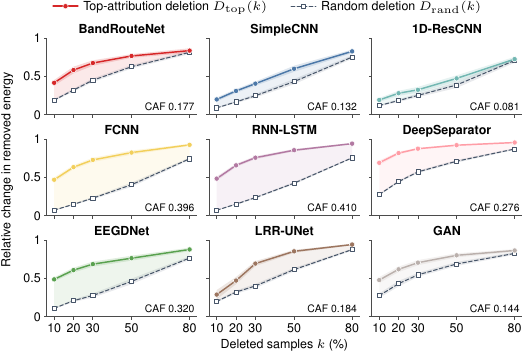}
    \caption{Visualization of CAF across five different fractions on EEGDenoiseNet-EOG dataset.}
    \label{fig:causal-eog}
    \vspace{-0.5cm}
    
\end{figure}

\textbf{CAF:}
CAF evaluates whether perturbing highly attributed
input samples produces a greater change in the
removed-energy target than perturbing randomly
selected samples. For an attribution method $\mathcal A$, let
$\Omega_{k,\mathcal A}^{\mathrm{top}}$ contain
the top-$k$ fraction of samples ranked by
$\mathbf{S}_{\mathcal A}(\mathbf{y})$.
Let $\mathbf{y}_{\setminus\Omega}$ denote the input
after zero-baseline deletion of the selected samples.
Using the removed-energy target in
Eq.(~\ref{qg:removed_energy_target}), we define
\begin{equation}
    D_{\mathrm{top},\mathcal A}(k)
    =
    \frac{
        \left|
            g(\mathbf{y})
            -
            g\left(
                \mathbf{y}_{\setminus
                \Omega_{k,\mathcal A}^{\mathrm{top}}}
            \right)
        \right|
    }{
        |g(\mathbf{y})|+\epsilon
    }.
    \label{eq:dtop}
\end{equation}
To assess whether highly attributed samples have a greater influence on the model's removed-energy response than randomly selected samples, we also compute random-deletion baseline
$D_{\mathrm{rand}}(k)$ as the mean normalized
response obtained by deleting randomly selected samples at the same fraction $k$.

Then, the attribution-specific score is calculated as
\begin{equation}
    \mathrm{CAF}_{\mathcal A}(k)
    =
    \operatorname{clip}
    \left(
        D_{\mathrm{top},\mathcal A}(k)
        -D_{\mathrm{rand}}(k),
        0,1
    \right),
    \label{eq:caf}
\end{equation}
and is averaged across the evaluated deletion fractions:
\begin{equation}
    \mathrm{CAF}
    =
    \frac{1}{|\mathcal K|}
    \sum_{k\in\mathcal K}
    \mathrm{CAF}_{\mathcal A}(k).
    \label{eq:caf_aggregate}
\end{equation}

We use five values of top-$k$ fractions (in Table~\ref{tab:trix_configuration})  and average three random zero deletion baselines for each fraction. An example of the comparison between top-$k$ and random zero-baseline deletions across model is shown at Fig.~\ref{fig:causal-eog}.

\begin{figure}[t]
    \centering
    \includegraphics{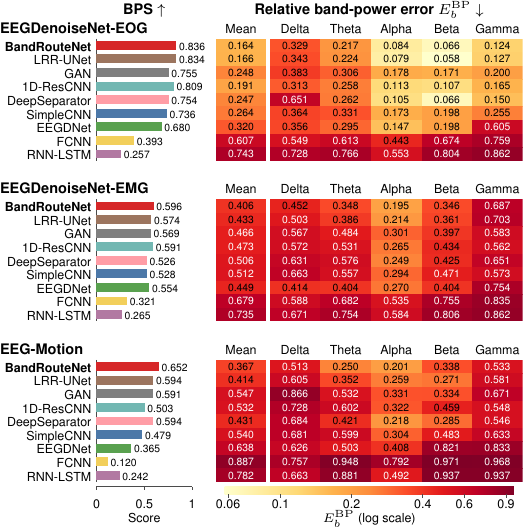}
    \caption{SV across datasets. Left: BPS (higher is better). Right: relative band-power error
    $E_b^{\mathrm{BP}}$ per band and its mean $\overline{E}^{\mathrm{BP}}$
    (lower is better); colours use a log scale shared by all datasets.}
    \label{fig:bandpower}
    \vspace{-0.3cm}
\end{figure}

\textbf{ASS:}
ASS assesses the consistency of the attribution map
under a small input perturbation.
Let
$\widetilde{\mathbf{y}}
=\mathbf{y}+\boldsymbol{\delta}$,
where $\boldsymbol{\delta}$ denotes a perturbation
chosen to produce limited changes to the input signal.
For a fixed attribution method $\mathcal A$, we define
\begin{equation}
    \mathrm{ASS}
    =
    1-
    \frac{
        \left\|
            \mathbf{S}_{\mathcal A}(\mathbf{y})
            -
            \mathbf{S}_{\mathcal A}
            (\widetilde{\mathbf{y}})
        \right\|_1
    }{
        \left\|
            \mathbf{S}_{\mathcal A}(\mathbf{y})
        \right\|_1
        +
        \left\|
            \mathbf{S}_{\mathcal A}
            (\widetilde{\mathbf{y}})
        \right\|_1
        +\epsilon
    }.
    \label{eq:ass}
\end{equation}

The score is bounded in $[0,1]$, with higher values
indicating greater consistency under the specified
perturbation

The Level-3 Reliability Score (RS) is calculated as
\begin{equation}
    \mathrm{RS}=\frac{1}{4}\left(
    \mathrm{TAAS}+\mathrm{BAC}+\mathrm{CAF}+\mathrm{ASS}\right).
    \label{eq:rs}
\end{equation}



\subsection{TriX-EEG Final Score}
\label{subsec:tds}

The overall TDS uses an equal-weight geometric mean,
\begin{equation}
    \mathrm{TDS}=(\mathrm{DS}\,\mathrm{FS}\,\mathrm{RS})^{1/3}.
    \label{eq:tds}
\end{equation}

The geometric mean aims to limit compensation: a high score at one level cannot fully
mask a failure at another. Nevertheless, TDS is not intended to replace the
level-wise analysis. We report DS, FS, and RS alongside TDS because they reveal
whether a limitation arises from SV, RF, or ER.

\section{Experimental Settings}
\label{sec:experimental_settings}

\subsection{Datasets}
\label{subsec:datasets}

We evaluate \textit{TriX-EEG} on two semi-simulated benchmarks derived from
EEGDenoiseNet~\cite{eegdenoisenet} and one EEG paired real-motion
benchmark from PhysioNet~\cite{motion_dataset}, referred to as \textit{EEG-Motion}. 
Two additional datasets are leveraged for external
validation beyond the primary benchmarks: EEG/EOG dataset~\cite{semi-eeg-eog} measures transfer to an
unseen denoising domain, and BCI Competition IV-2a~\cite{bci_iv_2a} measures
preservation of task-relevant motor-imagery information.

\subsubsection{EEGDenoiseNet}
The dataset contains 4515 clean EEG segments, 3400 EOG artifact segments, and 5598 EMG artifact segments. Each segment is 2-second length and is sampled at 256 Hz.
Following the original construction protocol~\cite{eegdenoisenet}, clean EEG segments are first divided into training, validation, and test partitions in an 8:1:1 ratio. EOG or EMG artifacts are then mixed with the clean segments within each partition at ten SNR levels ranging from $-7$ to $2$~dB, producing two artifact-specific benchmarks, denoted \textit{EEGDenoiseNet-EOG} and \textit{EEGDenoiseNet-EMG}. The resulting clean--artifact pairings are kept fixed across denoisers, ensuring consistent data partitions and enabling paired model comparisons.

\subsubsection{EEG-Motion}
The \textit{EEG-Motion} benchmark is constructed from the PhysioNet
motion-artifact dataset~\cite{motion_dataset}, which provides simultaneously
recorded clean and motion-contaminated EEG channels.
The complete recordings undergo polynomial baseline correction~\cite{motion-baseline-cr}, a
fourth-order 0.1--100~Hz band-pass filter, powerline noise notch filters, and
resampling from 2,048 to 256~Hz before splitting into 2-s windows. The clean and
contaminated channels are processed independently, and the reference artifact
is defined sample-wise as their difference. Trial identifiers are split before
window extraction to prevent trial leakage. The data are also split 8:1:1 into training, validation, and test sets.

\subsubsection{EEG/EOG dataset}
The dataset includes 19-channel EEG and corresponding EOG recordings from 27 subjects, sampled at 200~Hz and segmented into 30-s trials. Semi-simulated contaminated EEG is generated by linearly mixing clean EEG with real EOG recordings following~\cite{semi-eeg-eog}. We select nine EOG-sensitive frontal and temporal channels (Fp1, Fp2, F3, F4, Fz, F7, F8, T3, and T4), resample the signals to 256~Hz, apply a 1--80~Hz bandpass filter, and segment them into non-overlapping 2-s multichannel windows. This dataset serves exclusively as an unseen domain for assessing whether denoiser rankings established on EEGDenoiseNet-EOG generalize to a different paired EEG–EOG setting. To prevent data leakage, all subjects and windows are excluded from denoiser training, checkpoint selection, and hyperparameter tuning.

\subsubsection{BCI Competition IV-2a}

BCI Competition IV-2a~\cite{bci_iv_2a} comprises four-class motor-imagery EEG
(left hand, right hand, both feet, and tongue) from nine subjects, recorded in
two sessions on different days. Each recording contains 22 EEG and three EOG
channels sampled at 250~Hz, together with expert trial-rejection annotations.
For each trial, we extract the 0.5--2.5~s interval after cue onset and resample it to 256~Hz,
yielding one 512-sample trial window. 
The labeled calibration session (session~T) is used exclusively to fit the
downstream classifiers, whereas the independent evaluation session
(session~E) is used only for final testing. After removal of trials carrying
the expert artifact marker, session~T contains 2,328 trials and session~E contains 2,368 trials.

\subsection{Implementation Details}
\label{subsec:implementation_details}

\subsubsection{Denoisers and Training}
We evaluate \textit{TriX-EEG} using nine trained EEG denoisers selected to represent a
broad range of architectural designs. The evaluated models are
SimpleCNN~\cite{eegdenoisenet},
1D-ResCNN~\cite{1D-ResCNN},
FCNN~\cite{eegdenoisenet},
RNN-LSTM~\cite{eegdenoisenet},
DeepSeparator~\cite{deepseparator},
EEGDNet~\cite{EEGDnet},
LRR-UNet~\cite{lrr_unet},
GAN~\cite{denoise_gan}, and
BandRouteNet~\cite{brn}.

All evaluated models were implemented in PyTorch and trained on an NVIDIA Tesla T4 GPU using the AdamW optimizer~\cite{adamw} with an initial learning rate of $1\times10^{-3}$, weight decay of $1\times10^{-4}$. Training was conducted for 15 epochs on EEGDenoiseNet-EOG and EEGDenoiseNet-EMG, and 100 epochs on EEG-Motion.

\subsubsection{TriX-EEG Evaluation}
\label{subsubsec:trix_evaluation_protocol}

Levels~1 and~2 depend only on the fixed model output and paired signal
references. They are therefore deterministic for a given test set and model
checkpoint. Both levels are evaluated once over the complete test set. Component metrics are first calculated for each EEG window
and subsequently macro-averaged. Because no stochastic
operation is involved, Level-1 and Level-2 results are reported as single
values rather than as mean and standard deviation.

We evaluate Level~3 on five independently sampled, stratified representative subsets, owing to the computational cost. 
Each subset
contains 64 test windows that do not overlap with other subsets. SmoothGrad~\cite{smoothgrad} is leveraged to generate model attribution. Within a given dataset and replicate, all models receive
the same window indices and stochastic-operation seed. Level-3 metrics are reported as the mean and standard deviation over the
five subset/operation replicates. 
The normalization thresholds and
Level-3 operation settings are fixed before model comparison and are applied
consistently to every model and dataset. Table~\ref{tab:trix_configuration}
summarizes the complete configuration.

\begin{table}[t]
\centering
\caption{Configuration of the TriX-EEG evaluation framework.}
\label{tab:trix_configuration}
\footnotesize
\renewcommand{\arraystretch}{1.12}
\setlength{\tabcolsep}{3pt}
\scalebox{0.85}{
\begin{tabularx}{\columnwidth}{
@{}p{0.43\columnwidth}
p{0.18\columnwidth}
X@{}
}
\toprule
\textbf{Parameter} &
\textbf{Metric(s)} &
\textbf{Configuration} \\
\midrule

\multicolumn{3}{@{}l}{\textit{Level 1: Signal Validity}} \\
\addlinespace[2pt]

Maximum SNR improvement
& $U_{\Delta\mathrm{SNR}}$
& $\Delta\mathrm{SNR}_{\max}=20$ dB \\

Band-power error tolerance
& BPS
& $E_{\max}^{\mathrm{BP}}=1$ \\
\midrule
\addlinespace[3pt]
\multicolumn{3}{@{}l}{\textit{Level 2: Removal Faithfulness}} \\
\addlinespace[2pt]

Removed-energy deviation tolerance
& $U_{\mathrm{RAER}}$
& $E_{\max}^{\mathrm{RAER}}=1$ \\

ARE tolerance
& $U_{\mathrm{ARE}}$
& $E_{\max}^{\mathrm{ARE}}=1$ \\

AGE tolerance
& $U_{\mathrm{AGE}}$
& $E_{\max}^{\mathrm{AGE}}=1$ \\

CNL tolerance
& $U_{\mathrm{CNL}}$
& $E_{\max}^{\mathrm{CNL}}=1$ \\

Ridge fraction
& AGE, CNL
& $\lambda = 10^{-6}$ of the mean Gram-matrix diagonal \\
\midrule
\addlinespace[3pt]
\multicolumn{3}{@{}l}{\textit{Level 3: Explanation Reliability}} \\
\addlinespace[2pt]

STFT window
& TAAS
& Hann, 128 samples \\

STFT hop length
& TAAS
& 32 samples \\

Artifact frequency range
& TAAS
& $\mathcal{F}=[0.5,45]$ Hz \\

Deletion fractions
& CAF
& 10\%, 20\%, 30\%, 50\% and 80\% \\

Deletion baseline
& CAF
& Zero masking \\

Random deletion controls
& CAF
& Three per deletion fraction \\

Perturbation distribution
& ASS
& Zero-mean Gaussian \\

Perturbation magnitude
& ASS
& 1\% of input-window root-mean-square (RMS) \\

Subset size
& Level 3
& 64 windows \\

Subset/operation repetitions
& RS, TDS
& Five seeds: 42--46 \\
\midrule
\addlinespace[3pt]
\multicolumn{3}{@{}l}{\textit{General Evaluation Settings}} \\
\addlinespace[2pt]

Within-level aggregation
& DS, FS, RS
& Equal-weight arithmetic mean \\

Cross-level aggregation
& TDS
& Equal-weight geometric mean \\

Dataset split seed
& All metrics
& 42 \\

\bottomrule
\end{tabularx}
}
\vspace{-0.3cm}
\end{table}

\begin{figure}[t]
    \centering
    \includegraphics{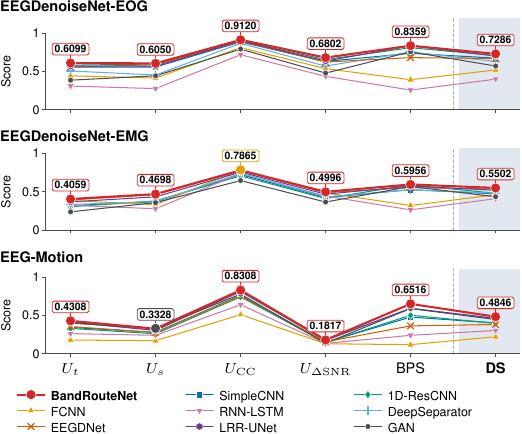}
    \caption{Level-1 metric profiles of denoising models across datasets, constructed by averaging per-window clipped utility scores ($U_t$, $U_s$: temporal and spectral RRMSE utilities; $U_{\mathrm{CC}}$: correlation utility). The shaded final axis presents the aggregate DS. The annotations highlight the best-performing value for each metric component.}
    \label{fig:lv1-metrics}
    \vspace{-0.3cm}
\end{figure}
\begin{figure}[t]
    \centering
    \includegraphics{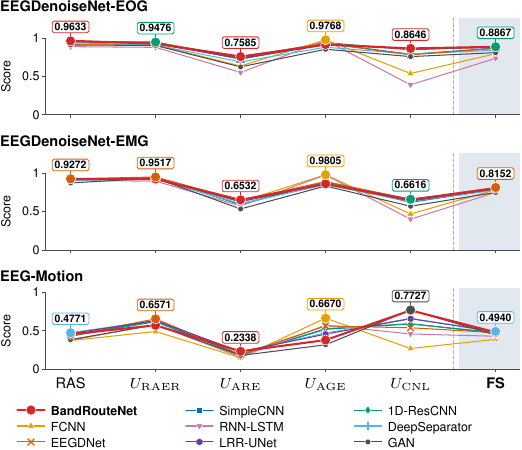}
    \caption{Level-2 RF profiles. The shaded axis gives the aggregate FS.}
    \label{fig:lv2-metrics}
    \vspace{-0.3cm}
    
\end{figure}
\begin{figure}[t]
    \centering
    \includegraphics{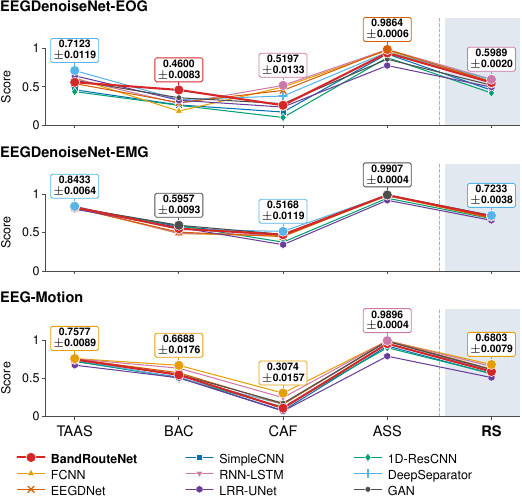}
    \caption{Level-3 ER profiles. Points show the mean over
    five stratified 64-window subsets, and error annotations report the sample
    standard deviation across subset/operation replicates. The shaded final axis
    reports the aggregate RS.}
    \label{fig:lv3-metrics}
    \vspace{-0.3cm}
    
\end{figure}


\begin{figure*}[t]
        \centering
    \includegraphics{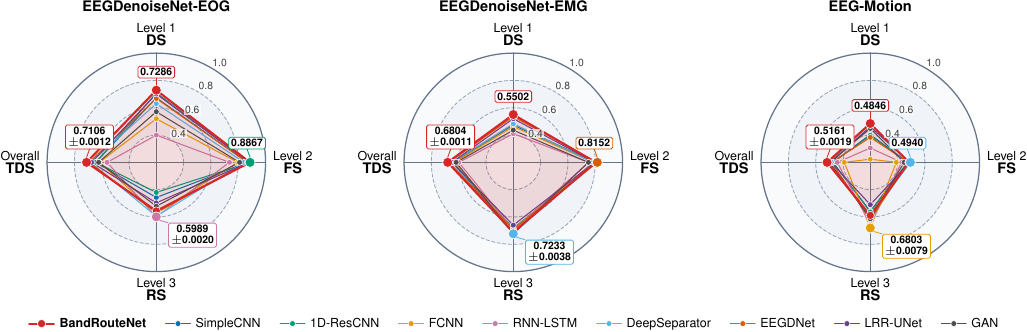}
    \caption{Level-wise metrics across datasets. Boxes give the best value on each axis; for RS and TDS, mean $\pm$ standard deviation over the five Level-3 subsets.}
    \label{fig:level-metrics}
    \vspace{-0.3cm}
\end{figure*}
    



\subsubsection{Cross-Dataset Validation Protocol}
\label{subsubsec:cross_dataset_protocol}

The nine EEGDenoiseNet-EOG checkpoints are applied independently to each of
the nine selected channels in the EEG/EOG dataset without retraining or fine-tuning. Each contaminated channel window is
divided by its own standard deviation without mean centering, matching the
EEGDenoiseNet checkpoint adapter. We form two target-domain orderings. The
first uses only conventional reconstruction measures; the second applies all
three \textit{TriX-EEG} levels. Levels~1 and~2 use all  paired windows,
whereas Level~3 uses five paired, approximately subject-balanced subsets of 64
single-channel windows to match the source-domain settings.

Rank transfer is evaluated against the artifact-matched EEGDenoiseNet-EOG
ordering using Spearman's $\rho$. Exact two-sided
permutation $p$-values are computed over the nine denoisers. Uncertainty in external TDS is summarized
over the five paired Level-3 replicates.

\subsubsection{Downstream Validation Protocol under Controlled EOG Contamination}
\label{subsubsec:downstream_protocol}
We investigate whether denoisers ranked highly by the artifact-matched \textit{TriX-EEG} evaluation also preserve task-relevant information and support downstream motor-imagery classification on BCI Competition IV-2a. Two settings are considered: (1)~\emph{single-channel}, where C3 and C4 are denoised and classified separately using subject-specific filter-bank log-power shrinkage-LDA classifiers~\cite{LDA}, consistent with the denoisers' single-channel training setup; and (2)~\emph{multichannel}, where all 22 EEG channels are denoised independently before classification using a subject-specific spatial FBCSP--LDA model~\cite{fbscp_lda}, assessing the transfer of single-channel denoisers to multichannel decoding. In both settings, classifiers and preprocessing statistics are fitted once on clean training session~T and then frozen. All nine denoisers use pretrained EEGDenoiseNet-EOG checkpoints without fine-tuning on BCI IV-2a.

For evaluation, each clean trial from session~E is contaminated with EOG artifacts at six power-based SNR levels, $\{-15,-10,-5,0,2,5\}$~dB, covering severe to mild contamination. For each target subject, EOG waveforms and regression-based EOG--EEG propagation maps are estimated exclusively from the other eight subjects, ensuring subject-disjoint contamination. Before denoising, signals are bandpass-filtered and normalized channel-wise using the contaminated-input standard deviation to match the denoiser's preprocessing. The same frozen classifier evaluates downstream performance on clean, contaminated, and denoised trials, allowing direct comparison of downstream performance across denoisers. All denoisers are tested on identical trials, EOG donors, SNR levels, and contamination realizations.

The primary endpoint is subject- and SNR-balanced macro-F1 recovery relative
to the paired contaminated input,
\begin{equation}
    \Delta\mathrm{F1}_{m}
    =\mathrm{F1}\!\left(\mathcal{C}(D_m(\widetilde{\mathbf{y}}))\right)
    -\mathrm{F1}\!\left(\mathcal{C}(\widetilde{\mathbf{y}})\right),
    \label{eq:downstream_recovery}
\end{equation}
where $\mathcal{C}$ is the frozen classifier, $D_m$ is denoiser $m$, and
$\widetilde{\mathbf{y}}$ is the contaminated trial. Positive values indicate
downstream recovery. Accuracy improvement is also analyzed analogously. Rank agreement with the
artifact-matched DS, FS, RS, and TDS scores is evaluated using exact two-sided
Spearman permutation tests over the nine denoisers.
 
\section{Results and Discussion}
\subsection{TriX-EEG Evaluation Results}
\subsubsection{Level 1: Signal Validity - Conventional Benchmark}
\label{subsec:level1_results}

Level~1 reflects the standard evaluation practice in the EEG-denoising
literature, where performance is commonly assessed using reconstruction error,
correlation, SNR improvement, and spectral preservation. It therefore serves
as both a conventional literature benchmark and the first level of our broader
evaluation framework. Raw metric values are reported in
Table~\ref{tab:benchmark_results} and Fig.~\ref{fig:bandpower}, while their
normalized, higher-is-better utilities are summarized in
Fig.~\ref{fig:lv1-metrics}.

BandRouteNet achieves the highest DS on all three datasets and the highest
cross-dataset mean DS of $0.5878$
(Table~\ref{tab:mean_trix_component_scores}). On EOG, it leads temporal and
spectral reconstruction, correlation, $\Delta$SNR, and BPS. On EMG, it retains
the best reconstruction errors, $\Delta$SNR, and BPS while remaining
competitive in correlation. This consistent component profile suggests that
its overall advantage does not arise from a single dominant metric. Motion artifacts are more challenging, indicated by the lower DS and
smaller SNR improvements across most models. BandRouteNet nevertheless
achieves the best temporal error, correlation, $\Delta$SNR, and BPS, whereas
LRR-UNet obtains the lowest spectral RRMSE.
BandRouteNet's BPS values of $0.836$, $0.596$, and $0.652$ on EOG, EMG, and
motion, respectively, further indicate comparatively its strong preservation of physiologically relevant EEG rhythms.

The remaining models exhibit less balanced behavior. EEGDNet and 1D-ResCNN
are competitive on the synthetic EOG and EMG datasets.
LRR-UNet is
comparatively robust across datasets, although it does not consistently lead
the temporal and noise-reduction metrics. FCNN and RNN-LSTM can achieve
moderate or high correlation in some settings, yet their larger reconstruction
errors and weaker BPS produce substantially lower DS values. These cases
demonstrate that a favorable result on one conventional metric does not
necessarily imply reliable signal reconstruction.

Overall, Level~1 identifies BandRouteNet as the most balanced model under the
conventional literature benchmark. However, signal similarity alone cannot
establish whether the correct artifact was removed or whether the model's
explanation is faithful. These complementary questions are addressed by
Levels~2 and~3.

\subsubsection{Level 2: Removal Faithfulness Analysis}
\label{subsec:level2_results}

Level~2 examines whether the model removes the correct artifact while
preserving neural activity. Figure~\ref{fig:lv2-metrics} reveals that no model
dominates every component, demonstrating the importance of jointly evaluating
artifact morphology, energy, reconstruction, artifact gain, and
conditional leakage.

On EOG, 1D-ResCNN obtains the highest FS ($0.8867$). BandRouteNet leads RAS, $U_{\mathrm{ARE}}$, and
$U_{\mathrm{CNL}}$, whereas 1D-ResCNN obtains the best removed-energy match.
On EMG, EEGDNet achieves the highest FS ($0.8152$) and RAS, while BandRouteNet
leads $U_{\mathrm{ARE}}$ and 1D-ResCNN leads $U_{\mathrm{CNL}}$. These profiles
show that a similar aggregate FS can arise from different removal behaviors,
which motivates retaining the component-level analysis.

Motion artifacts are substantially more difficult: the best FS decreases to
$0.4940$ for DeepSeparator. The low RAS and $U_{\mathrm{ARE}}$ values indicate that
reconstructing the complete motion-artifact waveform remains difficult.
DeepSeparator benefits from the highest RAS, EEGDNet obtains the closest
removed-energy match, and GAN leads $U_{\mathrm{CNL}}$; no model therefore
dominates the motion profile. Across all datasets, as shown in Table~\ref{tab:mean_trix_component_scores} BandRouteNet has the highest
mean FS ($0.7244$), narrowly ahead of LRR-UNet ($0.7209$), EEGDNet ($0.7200$),
and 1D-ResCNN ($0.7191$). The small differences reinforce the need to inspect
the component profiles rather than relying exclusively on FS. Notably, AGE should also be
interpreted jointly with CNL due to their dependence: an artifact coefficient close to one can appear
favorable even when the removed component contains substantial clean-aligned
activity.

\subsubsection{Level 3: Explanation Reliability Analysis}
\label{subsec:level3_results}

Level~3 evaluates whether the model's explanation is temporally aligned with
the artifact, consistent with its spectral dependence, causally relevant, and
stable. As shown in Fig.~\ref{fig:lv3-metrics}, Fig.~\ref{fig:saliency-eog}, and Fig.~\ref{fig:saliency-emg}, the leading model varies across
datasets and components. DeepSeparator achieves the strongest temporal
alignment on EOG and EMG and the highest EMG RS ($0.7233$), whereas RNN-LSTM
obtains the highest EOG RS ($0.5989$). On motion, FCNN leads TAAS, BAC, CAF,
and the aggregate RS ($0.6803$).

ASS is consistently stable for most models, indicating that their
attributions are robust under small input perturbations. However, the larger
variation in TAAS, BAC, and particularly CAF shows that stability alone does
not guarantee artifact alignment or causal relevance. Across datasets as shown in Table~\ref{tab:mean_trix_component_scores},
RNN-LSTM achieves the highest mean RS ($0.6502\pm0.0051$), followed by FCNN
($0.6462\pm0.0050$). 
These findings suggest that RS captures a distinct aspect of model explainability, with comparatively simpler models exhibiting higher explanation reliability despite weaker denoising performance. This further emphasizes that RS should be interpreted conditionally alongside DS and FS: reliable explanation is meaningful only when it describes a denoiser that is itself
valid and faithful.

\subsubsection{Overall TriX-EEG Performance}
\label{subsec:overall_results}

Table~\ref{tab:mean_trix_component_scores} summarizes performance across the
three evaluation levels. BandRouteNet achieves the highest mean TDS
($0.6357\pm0.0014$), followed by EEGDNet ($0.6151\pm0.0015$) and
DeepSeparator ($0.6142\pm0.0013$). Its advantage combines the highest mean DS
and FS with a competitive RS. This balanced profile is precisely what the
geometric aggregation is intended to reward.

The overall ranking also demonstrates how the three levels reflect different aspects of evaluation criteria and they should be 
interpreted together. For example, RNN-LSTM achieves the highest mean RS but ranks last
in TDS because of its weak DS and FS. 
Similarly, FCNN has the second-highest
RS but only the eighth-highest TDS. 
Conversely, EEGDNet combines a competitive mean FS, DS, and RS, resulting in
the second-highest overall score.
\textit{TriX-EEG} therefore prevents a strong result at one level from masking poor
signal reconstruction, unfaithful artifact removal, or unreliable
explanations.

\subsubsection{Correlation and Cross-Level Complementarity}
\label{subsec:metric_correlation}

\begin{table}[t]
    \centering
    \caption{Mean \textit{TriX-EEG} component scores across the three datasets. DS and FS are computed over the complete test sets, whereas RS and TDS are reported as the mean ± standard deviation, calculated using the root-mean-square of the per-dataset standard deviations. Models are ranked by their mean TDS.}
    \label{tab:mean_trix_component_scores}
    \small
    \setlength{\tabcolsep}{8pt}
    \scalebox{0.69}{
    \begin{tabular}{@{}r l c c c c@{}}
        \toprule
        \textbf{Rank} & \textbf{Model} & \textbf{Mean DS} & \textbf{Mean FS} & \textbf{Mean RS} & \textbf{Mean TDS} \\
        \midrule
        1 & BandRouteNet & $\mathbf{0.5878}$ & $\mathbf{0.7244}$ & $0.6185 \mathbin{\pm} 0.0045$ & $\mathbf{0.6357} \mathbin{\pm} 0.0014$ \\
        2 & EEGDNet & $0.5287$ & $0.7200$ & $0.6265 \mathbin{\pm} 0.0054$ & $0.6151 \mathbin{\pm} 0.0015$ \\
        3 & DeepSeparator & $0.5231$ & $0.7064$ & $0.6386 \mathbin{\pm} 0.0043$ & $0.6142 \mathbin{\pm} 0.0013$ \\
        4 & LRR-UNet & $0.5611$ & $0.7209$ & $0.5554 \mathbin{\pm} 0.0065$ & $0.6030 \mathbin{\pm} 0.0021$ \\
        5 & SimpleCNN & $0.5167$ & $0.7113$ & $0.5761 \mathbin{\pm} 0.0072$ & $0.5876 \mathbin{\pm} 0.0021$ \\
        6 & 1D-ResCNN & $0.5285$ & $0.7191$ & $0.5552 \mathbin{\pm} 0.0069$ & $0.5845 \mathbin{\pm} 0.0022$ \\
        7 & GAN & $0.4850$ & $0.6692$ & $0.6190 \mathbin{\pm} 0.0074$ & $0.5793 \mathbin{\pm} 0.0025$ \\
        8 & FCNN & $0.4017$ & $0.6580$ & $0.6462 \mathbin{\pm} 0.0050$ & $0.5460 \mathbin{\pm} 0.0011$ \\
        9 & RNN-LSTM & $0.3717$ & $0.6392$ & $\mathbf{0.6502} \mathbin{\pm} 0.0051$ & $0.5336 \mathbin{\pm} 0.0012$ \\
        \bottomrule
    \end{tabular}
    }
\end{table}
    
\begin{figure}[t]
    \centering
    \includegraphics{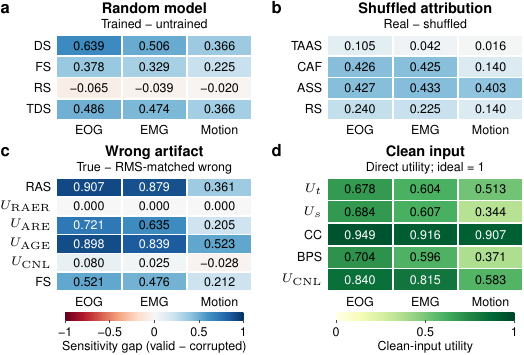}
    \caption{Sanity-check behavior of \textit{TriX-EEG}. Panels (a)--(c) report the
    valid-minus-corrupted sensitivity gap for trained versus untrained model,
    real versus shuffled attributions, and the true artifact
    versus an artifact of wrong type, respectively. Positive
    gaps indicate the expected response. Panel (d) reports higher-is-better
    utilities when clean EEG is supplied with a zero-artifact reference.}
    \label{fig:sanity_checks}
    \vspace{-0.3cm}
\end{figure}
\begin{table}[t]
    \centering
    \caption{Rank correlations among the three \textit{TriX-EEG} level scores. The
    dataset rows report Spearman correlations across nine models. The pooled
    row reports the correlation of normalized within-dataset ranks across all
    model--dataset evaluations.}
    \label{tab:level_score_correlations}
    \small
    \renewcommand{\arraystretch}{1.15}
    \setlength{\tabcolsep}{5pt}
    \scalebox{0.9}{
    \begin{tabular}{@{}lrrr@{}}
        \toprule
        \textbf{Analysis set}
        & \textbf{DS--FS}
        & \textbf{DS--RS}
        & \textbf{FS--RS} \\
        \midrule
        EEGDenoiseNet-EOG     & 0.9833 & -0.6500 & -0.7167 \\
        EEGDenoiseNet-EMG     & 0.8333 & -0.1667 & -0.4833 \\
        EEG-Motion            & 0.7833 & -0.7000 & -0.6167 \\
        \midrule
        Pooled within-dataset & 0.8667 & -0.5056 & -0.6056 \\
        \bottomrule
    \end{tabular}}
    \vspace{-0.3cm}
\end{table}

We analyze agreement among the three level-wise scores to assess whether they provide complementary information or simply induce similar model rankings. The nine models were ranked independently within each dataset for DS, FS, and RS; normalized ranks were then pooled across the three datasets to control for differences in dataset difficulty. We also computed dataset-wise Spearman correlations.

The rank-correlation analysis in Table~\ref{tab:level_score_correlations} reveals distinct relationships among the three evaluation levels. DS and FS exhibit strong pooled within-dataset rank agreement ($\rho=0.8667$), with correlations of $0.9833$, $0.8333$, and $0.7833$ for EOG, EMG, and motion artifacts, respectively. This agreement is consistent with the additive relationship in which reconstruction metrics at level 1 and removal errors in level 2 are mathematically coupled metrics. Thus, the strong DS–FS association is not interpreted as evidence of independent evaluation dimensions. Instead, their complementarity is primarily diagnostic: DS assesses the fidelity of the reconstructed EEG, whereas FS characterizes the removed component through reconstruction error and further quantifies conditionally artifact gain and clean-signal leakage. The differences in dataset-specific rankings and removal-component profiles indicate that reconstruction quality alone does not fully characterize how a denoiser separates neural activity from artifacts.

RS yields a markedly different model ordering. Its pooled correlations with DS
and FS are $-0.5056$ and $-0.6056$, respectively, and remain negative across
all three datasets. For example, 1D-ResCNN achieves the highest EOG FS
($0.8867$) but a relatively low RS (Fig.~\ref{fig:level-metrics}), whereas FCNN obtains the highest
motion RS ($0.6803$) despite the lowest DS and FS, as shown in Fig.~\ref{fig:lv1-metrics} - ~\ref{fig:lv3-metrics}. 
These observations indicate that stronger reconstruction and artifact-removal performance do not necessarily correspond to higher attribution reliability under the proposed evaluation criteria. However, the negative correlations are descriptive and should not be interpreted as evidence of a causal trade-off between denoising performance and explanation reliability. The disagreement involving RS could highlight a different assessment dimension. Thus, a high RS alone does not establish trustworthy denoising, as reliable attributions may still accompany poor signal reconstruction or inadequate artifact removal. 
Consequently, RS should be interpreted conditionally: Level~1 assesses reconstruction fidelity, Level~2 characterizes artifact removal and neural-signal leakage, while Level~3 evaluates the reliability of explanations associated with the denoising behavior.
Together, these findings support the complementary three-level design of \textit{TriX-EEG}, highlighting the need to assess denoising performance and ER jointly rather than relying on any single evaluation level.

\begin{table}[t]
    \centering
    \caption{Reconstruction-only target-domain performance on EEG/EOG dataset. Lower RRMSE and higher CC/$Q_{\mathrm{ext}}$ are better. The uncertainty of $Q_{\mathrm{ext}}$ is the sample standard deviation across subjects.}
    \label{tab:cross_eeg_external_ranking}
    \footnotesize
    \setlength{\tabcolsep}{3.2pt}
    \scalebox{0.87}{
    \begin{tabular}{@{}clrrrr@{}}
        \toprule
        \textbf{Rank} & \textbf{Denoiser} &
        $\mathbf{RRMSE}_{t}$ & $\mathbf{RRMSE}_{s}$
        & \textbf{CC} & $\mathbf{Q}_{\mathrm{ext}}$ \\
        \midrule
        \textbf{1} & \textbf{DeepSeparator} & $\mathbf{0.4009}$ & $0.4616$ & $\mathbf{0.9130}$ & $\mathbf{0.6835\pm0.0658}$ \\
        2 & LRR-UNet      & \underline{0.4208} & $\mathbf{0.4407}$ & \underline{0.8979} & \underline{$0.6789$}$\pm0.0758$ \\
        3 & BandRouteNet  & 0.4265 & \underline{0.4479} & 0.8970 & $0.6743\pm0.0722$ \\
        4 & GAN           & 0.4816 & 0.4749 & 0.8623 & $0.6365\pm0.0748$ \\
        5 & EEGDNet       & 0.4901 & 0.4833 & 0.8676 & $0.6314\pm0.0645$ \\
        6 & 1D-ResCNN     & 0.5882 & 0.5605 & 0.8049 & $0.5521\pm0.0673$ \\
        7 & SimpleCNN     & 0.5902 & 0.6048 & 0.8103 & $0.5385\pm0.0596$ \\
        8 & RNN-LSTM      & 0.8685 & 0.8837 & 0.5244 & $0.2574\pm0.0523$ \\
        9 & FCNN          & 0.8731 & 0.8951 & 0.5171 & $0.2496\pm0.0490$ \\
        \bottomrule
    \end{tabular}}
\end{table}

\begin{table}[t]
    \centering
    \caption{Complete target-domain \textit{TriX-EEG} evaluation on EEG-EOG dataset. RS and TDS are mean $\pm$ sample standard deviation over five paired Level-3 subsets.}
    \label{tab:cross_eeg_external_tds}
    \footnotesize
    \setlength{\tabcolsep}{3.8pt}
    \scalebox{0.82}{
    \begin{tabular}{@{}clrrrr@{}}
        \toprule
        \textbf{Rank} & \textbf{Denoiser} &
        \textbf{DS} & \textbf{FS} & \textbf{RS} & \textbf{TDS} \\
        \midrule
        \textbf{1} & \textbf{DeepSeparator} &
        $0.5979$ & $\mathbf{0.5398}$ &
        $0.5127\pm0.0075$ & $\mathbf{0.5490\pm0.0027}$ \\
        2 & BandRouteNet  & 0.5867 & 0.5251 & $0.4777\pm0.0097$ & $0.5280\pm0.0036$ \\
        3 & LRR-UNet      & $\mathbf{0.5980}$ & 0.5334 & $0.4470\pm0.0118$ & $0.5224\pm0.0046$ \\
        4 & EEGDNet       & 0.5294 & 0.4987 & $0.4936\pm0.0136$ & $0.5070\pm0.0046$ \\
        5 & GAN           & 0.5484 & 0.4861 & $0.4724\pm0.0076$ & $0.5012\pm0.0027$ \\
        6 & 1D-ResCNN     & 0.4645 & 0.4714 & $0.4257\pm0.0090$ & $0.4534\pm0.0032$ \\
        7 & SimpleCNN     & 0.4377 & 0.4625 & $0.4478\pm0.0096$ & $0.4492\pm0.0032$ \\
        8 & RNN-LSTM      & 0.1830 & 0.3586 & $\mathbf{0.5466\pm0.0094}$ & $0.3298\pm0.0019$ \\
        9 & FCNN          & 0.1783 & 0.3582 & $0.5219\pm0.0098$ & $0.3218\pm0.0020$ \\
        \bottomrule
    \end{tabular}}
\end{table}
\subsubsection{Sanity-Check Validation}
\label{subsec:sanity_checks}
We conduct a set of targeted sanity checks to examine whether the proposed
metrics respond appropriately to controlled failure modes. Specifically, we
evaluate an untrained model, shuffled attribution maps, artifact references of
the wrong type, and clean inputs with zero-artifact references. 
Additional analyses, including
explanation-method sensitivity, normalization-threshold sensitivity, are provided in the appendix. EEGDNet~\cite{EEGDnet} is used as the representative model for these experiments.

For controlled corruptions, we report the sensitivity gap
\begin{equation}
    G=S_{\mathrm{valid}}-S_{\mathrm{corrupted}},
    \label{eq:sanity_gap}
\end{equation}
with all scores oriented so that higher is better. Thus, $G>0$ indicates the
expected response, whereas $G<0$ indicates counterintuitive behavior.
Clean-input preservation utilities are reported directly, with an ideal value
of one.

\paragraph{Untrained Model}
Randomizing the model parameters produced a clear end-to-end failure: DS
decreased by $0.366$--$0.639$, FS by $0.225$--$0.378$, and TDS by
$0.366$–$0.486$ (Fig.~\ref{fig:sanity_checks}a). However, RS increased
slightly in this case. This is because RS describes the explanation of a model's behavior, not the clinical or signal-processing validity of that behavior. The result reinforces the framework's hierarchical interpretation: RS should be interpreted only after DS and FS
establish denoising validity.

\paragraph{Attribution Corruption}
Shuffling the attribution maps reduced every attribution-dependent metric on
all datasets (Fig.~\ref{fig:sanity_checks}b). CAF and ASS showed the largest
and most consistent gaps, at $0.140$--$0.426$ and $0.403$--$0.433$,
respectively. These results support the complementary design of Level~3:
temporal alignment alone is insufficient, while causal relevance and stability
provide stronger evidence that an explanation reflects the model's behavior.

\paragraph{Wrong-Artifact Interference}
We replace the true artifact with a randomly paired, RMS-matched artifact from
another type. As shown in Fig.~\ref{fig:sanity_checks}c, wrong pairing sharply
reduces RAS, $U_{\mathrm{ARE}}$, and $U_{\mathrm{AGE}}$. Because artifact
energy is controlled via RMS-matched amplitude normalization, $U_{\mathrm{RAER}}$ is unchanged; the response of the
other metrics therefore demonstrates sensitivity to artifact identity rather
than a trivial amplitude mismatch. FS decreases by $0.521$, $0.476$, and
$0.212$ on EOG, EMG, and motion, respectively. The large reduction in $U_{\mathrm{AGE}}$ indicates
that the removed component no longer exhibits the unit artifact gain expected
for the correct reference. CNL need not change
monotonically, possibly because replacing the artifact regressor can redistribute
variance between correlated clean and artifact predictors. Accordingly, the
joint component pattern should accompany the aggregate FS.
\begin{table}[t]
    \centering
    \caption{Rank agreement between the target-domain orderings and the source-domain \textit{TriX-EEG} scores on EEGDenoiseNet-EOG. Exact $p$-values are two-sided permutation tests across nine denoisers.}
    \label{tab:cross_eeg_rank_correlation}
    \footnotesize
    \setlength{\tabcolsep}{4pt}
    \scalebox{0.95}{
    \begin{tabular}{@{}llrr@{}}
        \toprule
        \textbf{Target ordering} & \textbf{Source ranking} &
        \textbf{Spearman $\boldsymbol{\rho}$} & \textbf{Exact $\boldsymbol{p}$} \\
        \midrule
        $Q_{\mathrm{ext}}$ & DS &
        $0.4500$ & $0.2298$ \\
        $Q_{\mathrm{ext}}$ & FS  & 0.4000 & 0.2912 \\
        $Q_{\mathrm{ext}}$ & TDS & \textbf{0.7000} & \textbf{0.0433} \\
        \addlinespace[2pt]
        $\mathrm{TDS}_{\mathrm{ext}}$ & DS  & 0.5167 & 0.1618 \\
        $\mathrm{TDS}_{\mathrm{ext}}$ & FS  & 0.4500 & 0.2298 \\
        $\mathrm{TDS}_{\mathrm{ext}}$ & RS  & 0.0667 & 0.8801 \\
        $\mathrm{TDS}_{\mathrm{ext}}$ & TDS &
        $\mathbf{0.8333}$ & $\mathbf{0.0083}$ \\
        \bottomrule
    \end{tabular}}
        \vspace{-0.3cm}

\end{table}
\paragraph{Clean Input with No Artifact}
The clean-input check reveals a separate limitation
(Fig.~\ref{fig:sanity_checks}d). Correlation remained high
($0.907$--$0.949$), but the temporal utility was only $0.513$--$0.678$, and the
spectral utility fell to $0.344$ on the motion data. In other words, the output
can retain the broad waveform shape while still changing a substantial amount
of clean activity. The clean-leakage utility $U_{\mathrm{CNL}}$ ranges from
$0.583$ on motion to $0.840$ on EOG, providing a complementary Level-2 warning
after controlling for the artifact contribution. This result identifies a lack
of identity preservation in the evaluated denoiser that conventional
correlation alone would understate.

\subsection{External Validation}
The following experiments investigate two distinct forms of validation: (1) Unseen-domain validation on EEG/EOG dataset~\cite{semi-eeg-eog} asks
whether the benchmark preserves a broad denoiser ordering under a dataset
shift, and (2) downstream task validation on BCI Competition IV-2a dataset~\cite{bci_iv_2a} verifies whether that ordering is associated
with preservation of task-relevant neural information. 

\subsubsection{Unseen-Domain Validation on EEG/EOG}
\label{subsec:cross_eeg_validation} Following the protocol in section~\ref{subsubsec:cross_dataset_protocol},
we examine whether the denoiser ordering identified by
\textit{TriX-EEG} on EEGDenoiseNet-EOG is preserved in an unseen target
domain. We evaluate this question
using two complementary target-domain orderings. The first is derived only from
conventional reconstruction measures and therefore provides a criterion outside
the complete \textit{TriX-EEG} aggregation. The second is obtained by applying all three
\textit{TriX-EEG} levels to the target data.
\begin{table*}[t]
    \centering
    \caption{Single-channel downstream performance on C3 and C4. Metrics are
    balanced over subjects and SNRs; deltas are relative to the contaminated
    input. Bold denotes the best denoiser recovery and underlining the
    runner-up in each delta column.}
    \label{tab:downstream_singlechannel_performance}
    \scriptsize
    \renewcommand{\arraystretch}{1.06}
    \setlength{\tabcolsep}{4pt}
    \scalebox{1.1}{
    \begin{tabular}{@{}lrrrrrrrr@{}}
        \toprule
        & \multicolumn{4}{c}{\textbf{C3}} 
        & \multicolumn{4}{c}{\textbf{C4}} \\
        \cmidrule(lr){2-5}\cmidrule(l){6-9}
        \textbf{Input/method} 
        & \textbf{Acc.$\uparrow$} 
        & \textbf{$\Delta$Acc.$\uparrow$} 
        & \textbf{F1$\uparrow$} 
        & \textbf{$\Delta$F1$\uparrow$}
        & \textbf{Acc.$\uparrow$} 
        & \textbf{$\Delta$Acc.$\uparrow$} 
        & \textbf{F1$\uparrow$} 
        & \textbf{$\Delta$F1$\uparrow$} \\
        \midrule
        Clean EEG       & $0.3622$ & $+0.0402$ & $0.3470$ & $+0.0906$ & $0.3562$ & $+0.0395$ & $0.3279$ & $+0.0807$ \\
        Contaminated    & $0.3220$ & $0.0000$ & $0.2564$ & $0.0000$ & $0.3167$ & $0.0000$ & $0.2471$ & $0.0000$ \\
        \midrule
        \textbf{BandRouteNet} & $0.3311$ & $+0.0092$ & $0.2820$ & $\mathbf{+0.0256}$ & $0.3283$ & $\underline{+0.0116}$ & $0.2705$ & $\mathbf{+0.0234}$ \\
        LRR-UNet        & $0.3325$ & $\underline{+0.0105}$ & $0.2810$ & $\underline{+0.0246}$ & $0.3301$ & $\mathbf{+0.0134}$ & $0.2699$ & $\underline{+0.0228}$ \\
        DeepSeparator   & $0.3298$ & $+0.0078$ & $0.2734$ & $+0.0170$ & $0.3277$ & $+0.0110$ & $0.2644$ & $+0.0172$ \\
        EEGDNet         & $0.3348$ & $\mathbf{+0.0128}$ & $0.2729$ & $+0.0165$ & $0.3246$ & $+0.0079$ & $0.2604$ & $+0.0133$ \\
        1D-ResCNN       & $0.3237$ & $+0.0017$ & $0.2717$ & $+0.0153$ & $0.3260$ & $+0.0093$ & $0.2672$ & $+0.0201$ \\
        GAN             & $0.3209$ & $-0.0010$ & $0.2695$ & $+0.0131$ & $0.3185$ & $+0.0018$ & $0.2582$ & $+0.0111$ \\
        SimpleCNN       & $0.3104$ & $-0.0116$ & $0.2462$ & $-0.0102$ & $0.3140$ & $-0.0027$ & $0.2473$ & $+0.0002$ \\
        RNN-LSTM        & $0.2663$ & $-0.0556$ & $0.1498$ & $-0.1066$ & $0.2622$ & $-0.0544$ & $0.1386$ & $-0.1085$ \\
        FCNN            & $0.2602$ & $-0.0618$ & $0.1467$ & $-0.1097$ & $0.2624$ & $-0.0543$ & $0.1408$ & $-0.1063$ \\
        \bottomrule
    \end{tabular}
    }
\end{table*}

\paragraph{Reconstruction-Only External Criterion}
To avoid defining the first target ordering with the complete DS formula, we
combine only temporal RRMSE, spectral RRMSE, and waveform correlation. For
window $i$, we define
\begin{equation}
\begin{split}
Q_{\mathrm{ext}}^{(i)}=\frac{1}{3}\big[&
1-\operatorname{clip}(\mathrm{RRMSE}_{t}^{(i)},0,1) \\
&+1-\operatorname{clip}(\mathrm{RRMSE}_{s}^{(i)},0,1)
+\operatorname{clip}(\mathrm{CC}^{(i)},0,1)\big].
\end{split}
\label{eq:cross_eeg_external_score}
\end{equation}
The score is computed per window, macro-averaged within each subject, and then
summarized across subjects.

Table~\ref{tab:cross_eeg_external_ranking} shows a clear separation between
the leading and trailing methods. DeepSeparator, LRR-UNet, and BandRouteNet
form the top group and remain competitive across all three reconstruction
quantities. In contrast, RNN-LSTM and FCNN occupy the last two positions, with
substantially larger temporal and spectral errors. 

\paragraph{Complete Target-Domain TriX-EEG Evaluation}

We next apply the complete framework to EEG/EOG dataset, following the same protocol as the source-domain settings on EEGDenoiseNet-EOG. Sampling is approximately balanced across
subjects, and every denoiser receives identical subset indices and
operation seeds. The target-domain \textit{TriX-EEG} score is
\begin{equation}
\mathrm{TDS}_{\mathrm{ext}}=
\left(
\mathrm{DS}_{\mathrm{ext}}
\mathrm{FS}_{\mathrm{ext}}
\mathrm{RS}_{\mathrm{ext}}
\right)^{1/3}.
\label{eq:cross_eeg_external_tds}
\end{equation}

As shown in Table~\ref{tab:cross_eeg_external_tds}, DeepSeparator ranks first,
followed by BandRouteNet and LRR-UNet. The complete target-domain evaluation
therefore preserves the same top-three set found by the reconstruction-only
criterion, with only a swap between the second and third positions. The
ordering is also stable across Level-3 sampling: all five replicates produce
the same nine-model TDS order, and the TDS standard deviations range only from
0.0019 to 0.0046. Again, RS provides a counter-example on RS score (RNN-LSTM obtains the highest target-domain RS, but ranks eighth in TDS), indicating that ER
cannot be interpreted alone without denoising performance. 

\paragraph{Source-to-Target Rank Agreement}

\begin{table}[t]
    \centering
    \caption{Multichannel downstream performance with 22-channel FBCSP--LDA classifier.
    Metrics are balanced over subjects and SNRs; deltas are relative to the
    contaminated input. Bold denotes the best denoiser preservation and
    underlining the runner-up in each delta column.}
    \label{tab:downstream_multichannel_performance}
    \footnotesize
    \renewcommand{\arraystretch}{1.06}
    \setlength{\tabcolsep}{3.5pt}
    \begin{tabular}{@{}lrrrr@{}}
        \toprule
        \textbf{Input/method} & \textbf{Acc.}$\uparrow$ &
        $\Delta$\textbf{Acc.}$\uparrow$ &
        \textbf{Macro-F1}$\uparrow$ &
        $\Delta$\textbf{F1}$\uparrow$ \\
        \midrule
        Clean EEG       & $0.6440$ & $+0.0653$ & $0.6398$ & $+0.0761$ \\
        Contaminated    & $0.5787$ & $0.0000$ & $0.5637$ & $0.0000$ \\
        \midrule
        \textbf{BandRouteNet} & $0.5702$ & $\underline{-0.0085}$ & $0.5587$ & $\mathbf{-0.0050}$ \\
        DeepSeparator   & $0.5720$ & $\mathbf{-0.0067}$ & $0.5580$ & $\underline{-0.0057}$ \\
        LRR-UNet        & $0.5558$ & $-0.0229$ & $0.5418$ & $-0.0219$ \\
        EEGDNet         & $0.5532$ & $-0.0255$ & $0.5349$ & $-0.0288$ \\
        1D-ResCNN       & $0.5289$ & $-0.0498$ & $0.5045$ & $-0.0592$ \\
        SimpleCNN       & $0.5291$ & $-0.0496$ & $0.5027$ & $-0.0610$ \\
        GAN             & $0.4706$ & $-0.1081$ & $0.4497$ & $-0.1140$ \\
        RNN-LSTM        & $0.3913$ & $-0.1874$ & $0.3630$ & $-0.2007$ \\
        FCNN            & $0.3733$ & $-0.2054$ & $0.3423$ & $-0.2214$ \\
        \bottomrule
    \end{tabular}
        \vspace{-0.3cm}

\end{table}



Table~\ref{tab:cross_eeg_rank_correlation} shows that source-domain TDS exhibits the strongest correlation with both target-domain orderings. In particular, its correlation with external TDS reaches $\rho=0.8333$ (exact $p=0.0083$), exceeding those of source-domain DS ($\rho=0.5167$) and FS ($\rho=0.4500$). Combined with the consistent external TDS ordering across five paired subsets, these findings suggest that \textit{TriX-EEG} captures a broadly reproducible ranking of denoiser quality across datasets. Although exact rank invariance or universal generalization cannot be established, the agreement across target-domain orderings, the preservation of leading and trailing model groups, and the strong source-to-target TDS correlation supports the transferability of broad model rankings across the two evaluated EOG settings.

\begin{figure*}[t]
    \centering
    \includegraphics[width = \linewidth, height = 0.42\linewidth]{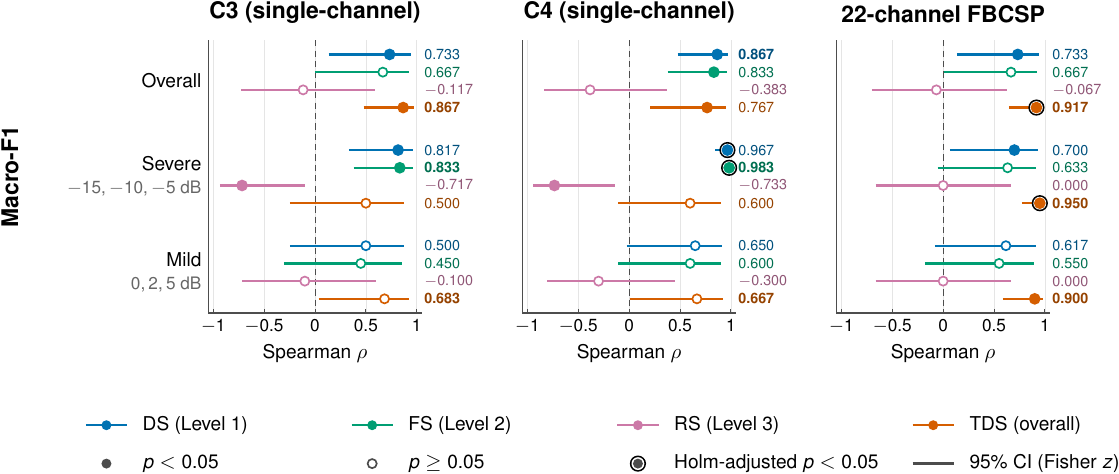}
    \caption{Agreement between the artifact-matched denoiser ranking produced by
    \textit{TriX-EEG} and the ranking by downstream macro-F1 recovery on
    BCI~IV-2a motor imagery ($n=9$ denoisers). Each row shows the Spearman
    correlation $\rho$ between one \textit{TriX-EEG} score (DS, FS, RS or TDS) and
    downstream recovery. Results are given over all SNR levels (Overall), severe
    contamination ($\{-15,-10,-5\}$~dB) and mild contamination ($\{0,2,5\}$~dB).
    Horizontal bars show approximate 95\% confidence intervals from the
    Fisher $z$-transform. Bold values mark the strongest association within each
    regime.}
\label{fig:downstream_agreement_f1}
    \vspace{-0.3cm}
\end{figure*}

\subsubsection{Downstream Task Validation under Controlled EOG Contamination}
\label{subsec:downstream_validation}
Following the protocol in Section~\ref{subsubsec:downstream_protocol}, we test whether denoisers ranked highly by the artifact-matched
\textit{TriX-EEG} evaluation also preserve information required for
motor-imagery classification. We consider two cases: (1)~\emph{single-channel},
in which denoising and classification are performed separately on C3 and C4;
and (2)~\emph{multichannel}, in which all 22 channels are denoised independently
before classification. The first case aligns with the single-channel setting in the source domain in which  
evaluated denoisers were trained, whereas the second additionally tests whether
channel-wise denoising preserves spatial information.

\paragraph{Downstream Recovery}

We first examine the channel-aligned case. As shown in Table~\ref{tab:downstream_singlechannel_performance}, six denoisers improve
C3 macro-F1 when averaged across SNRs. BandRouteNet ranks first
($\Delta\mathrm{F1}=+0.0256$), followed by
LRR-UNet ($+0.0246$) and DeepSeparator ($+0.0170$). C4 provides a consistent
replication: seven denoisers improve the contaminated input, led by
BandRouteNet ($+0.0234$) and LRR-UNet ($+0.0228$).
These gains show that denoising can recover downstream task-relevant information when the
denoiser and classifier operate on compatible single-channel representations. No method, however, restores performance to the clean reference.

The multichannel result is less favorable
(Table~\ref{tab:downstream_multichannel_performance}). Averaged over all SNRs,
no denoiser improves FBCSP macro-F1: BandRouteNet has the smallest reduction
($\Delta\mathrm{F1}=-0.0050$), followed by DeepSeparator ($-0.0057$). One plausible explanation is a mismatch between the denoisers' channel-independent training objective and the spatial covariance structure required by FBCSP–LDA. The EEGDenoiseNet-trained denoisers are applied independently to all 22
electrodes, without an objective that preserves inter-channel covariance or
relative amplitude, whereas FBCSP relies directly on those spatial relationships.

\paragraph{Agreement with \textit{TriX-EEG}}

First, the three downstream configurations exhibit highly consistent denoiser rankings, with pairwise Spearman correlations of $\rho=0.9333$ (C3–C4), $0.9667$ (C3–FBCSP), and $0.8833$ (C4–FBCSP). BandRouteNet consistently ranks first, while LRR-UNet and DeepSeparator remain among the leading methods. This consistency suggests that relative denoiser performance is largely preserved across electrodes and classification pipelines, despite differences in their absolute outcomes.

Figure~\ref{fig:downstream_agreement_f1} further examines whether the artifact-matched \textit{TriX-EEG} scores are associated with downstream model rankings. For single-channel C3, TDS exhibits the strongest overall agreement with macro-F1 ($\rho=0.867$). Positive associations are also observed for C4, although their magnitudes vary across evaluation levels and contamination regimes. Under severe contamination, DS and FS show particularly strong associations with downstream performance, with FS reaching $\rho=0.833$ for C3 and $0.983$ for C4 macro-F1. This pattern is consistent with reconstruction fidelity and artifact-removal quality becoming more consequential under substantial contamination. Under mild contamination, TDS generally provides the strongest or joint-strongest agreement, suggesting that its combined evaluation may better reflect relative performance when excessive signal alteration is a concern.

The multichannel FBCSP experiment provides a particularly informative test of ranking transfer. Although denoising does not improve average classification performance relative to the contaminated input, TDS maintains strong associations with downstream model rankings across both endpoints and all contamination regimes ($\rho=0.9000$–$0.9500$). The overall macro-F1 agreement ($\rho=0.9167$) indicates that the source-domain TDS ranking remains closely aligned with relative downstream performance even when the evaluated denoisers fail to provide an average classification benefit. Thus, the downstream evidence supports TDS as a correlate of relative performance preservation among the evaluated denoisers. In contrast, RS exhibits weak or negative associations with downstream recovery, indicating that the attribution properties measured by RS do not consistently track task-level performance, supporting the interpretation of RS as a complementary explanation-oriented assessment.

Overall, these results provide empirical evidence that the artifact-matched \textit{TriX-EEG} rankings, particularly TDS, are associated with relative downstream denoiser performance across different classification settings. The downstream experiments support the use of TDS for comparative denoiser assessment within the evaluated settings, while motivating the joint interpretation of all three levels rather than treating any individual score as a universal proxy for downstream utility.

\section{Limitations}

A primary limitation of \textit{TriX-EEG} is its reliance on ground-truth clean EEG and known artifact components, which are rarely available in real-world recordings. The composite score also depends on predefined normalization thresholds, aggregation weights, and attribution configurations. Finally, although external validation supports ranking consistency across the evaluated settings, generalization to other artifact types and clinical recordings remains uncertain. In practice, evaluation may depend on expert annotations or physiological plausibility measures, introducing uncertainty and bias. The findings may also vary across datasets, artifact types, models, and explanation methods. Future work should therefore develop reference-free evaluation and validate the framework on broader clinical and real-world EEG data.

\section{Conclusion}

This paper presented \textit{TriX-EEG}, a three-level framework for evaluating EEG artifact removal beyond conventional reconstruction metrics. By jointly assessing SV, RF, and ER, \textit{TriX-EEG} provides both an aggregate denoising score and a diagnostic profile of model behavior. Experiments across EOG, EMG, and motion-artifact benchmarks revealed complementary agreement between SV and RF, while ER exhibited distinct behavior capturing complementary aspects of model explainability. These results supports the joint interpretation of all three evaluation dimensions.
External validation demonstrated broad cross-domain ranking consistency, and downstream motor-imagery experiments showed that higher-ranked denoisers generally exhibited better relative performance preservation or introduced less performance degradation under controlled contamination. 
These findings demonstrate that reconstruction accuracy alone is insufficient to characterize trustworthy EEG artifact removal. Overall, \textit{TriX-EEG} offers a more comprehensive basis for comparing and diagnosing EEG denoisers beyond reconstruction accuracy alone, encouraging the development of models that are not only accurate, but also faithful and explainable.

\section{Acknowledgment}
This work was supported by Research NB - Student Support Fund No. SSF\_2027\_070. 


\begingroup
\renewcommand*{\bibfont}{\small}
\setlength\bibitemsep{0pt}
\setlength\bibparsep{0pt}
\printbibliography
\endgroup

\newpage
\input{appendix}

\end{document}

%% file: appendix.tex


\appendices

\section{Additional Model Comparisons}
\label{app:model_comparisons}

We complement the aggregate \textit{TriX-EEG}
scores with paired qualitative comparisons. For each dataset, we first form the same seed-42 evaluation set for Level 1-2 and five paired stratified 64-window subsets (seeds 42--46) for level 3. All models and explanation methods receive the same input within a dataset, which prevents sample selection from confounding the visual comparison. These examples illustrate model behavior and they do not replace the complete-test-set Level-1 and Level-2 evaluation or the five-subset Level-3 estimates.

\subsection{Signal and Removed-Artifact Reconstructions}

\begin{figure*}[t]
    \centering

    \subfloat[
        Level-1 qualitative comparison.
        The original noisy input and denoised output are
        overlaid on the paired clean EEG reference.
        \label{fig:app_l1_eog}
    ]{%
        \begin{minipage}[t]{0.49\textwidth}
            \centering
            \includegraphics{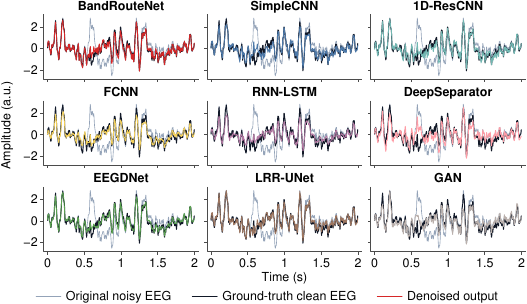}
        \end{minipage}%
    }
    \hfill
    \subfloat[
        Level-2 qualitative comparison.
        The original noisy input and the component removed
        by each model are overlaid on the ground-truth artifact.
        \label{fig:app_l2_eog}
    ]{%
        \begin{minipage}[t]{0.49\textwidth}
            \centering
            \includegraphics{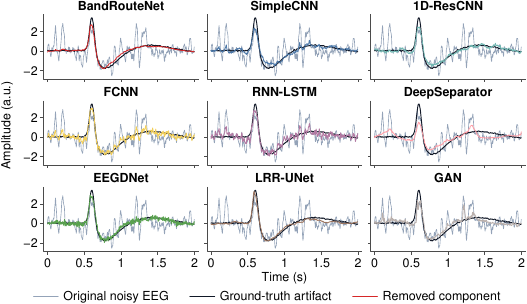}
        \end{minipage}%
    }

    \caption{Qualitative comparison on EEGDenoiseNet-EOG
    at two evaluation levels:
    (a) Level 1 assesses preservation of the clean EEG signal,
    while (b) Level 2 assesses whether the removed component
    corresponds to the ground-truth artifact.}
    \label{fig:app_eog_qualitative}
\end{figure*}

\begin{figure*}[t]
    \centering

    \subfloat[
        Level-1 qualitative comparison.
        The original noisy input and denoised output are
        overlaid on the paired clean EEG reference.
        \label{fig:app_l1_emg}
    ]{%
        \begin{minipage}[t]{0.49\textwidth}
            \centering
            \includegraphics{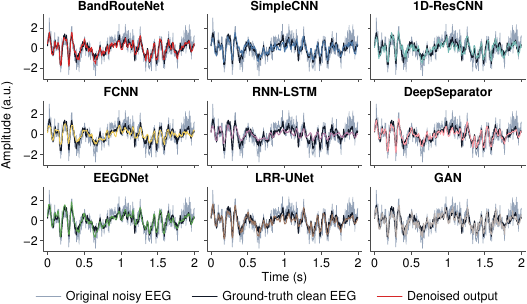}
        \end{minipage}%
    }
    \hfill
    \subfloat[
        Level-2 qualitative comparison.
        The original noisy input and the component removed
        by each model are overlaid on the ground-truth artifact.
        \label{fig:app_l2_emg}
    ]{%
        \begin{minipage}[t]{0.49\textwidth}
            \centering
            \includegraphics{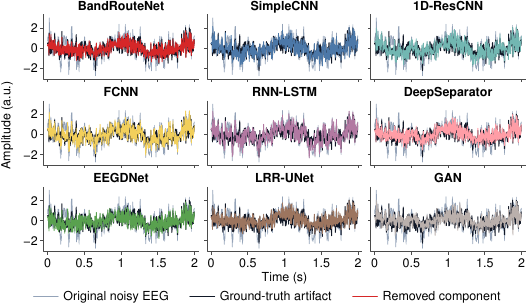}
        \end{minipage}%
    }

    \caption{Qualitative comparison on EEGDenoiseNet-EMG
    at two evaluation levels:
    (a) Level 1 assesses preservation of the clean EEG signal,
    while (b) Level 2 assesses whether the removed component
    corresponds to the ground-truth artifact.}
    \label{fig:app_emg_qualitative}
    \vspace{-0.3cm}
\end{figure*}

\begin{figure*}[t]
    \centering

    \subfloat[
        Level-1 qualitative comparison.
        The original noisy input and denoised output are
        overlaid on the paired clean EEG reference.
        \label{fig:app_l1_motion}
    ]{%
        \begin{minipage}[t]{0.49\textwidth}
            \centering
            \includegraphics{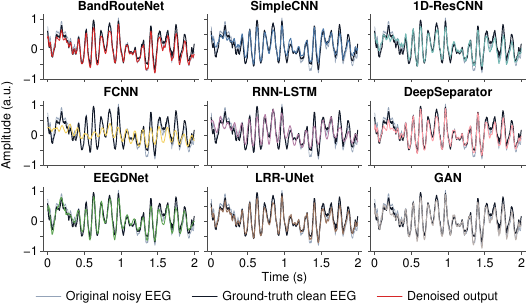}
        \end{minipage}%
    }
    \hfill
    \subfloat[
        Level-2 qualitative comparison.
        The original noisy input and the component removed
        by each model are overlaid on the ground-truth artifact.
        \label{fig:app_l2_motion}
    ]{%
        \begin{minipage}[t]{0.49\textwidth}
            \centering
            \includegraphics{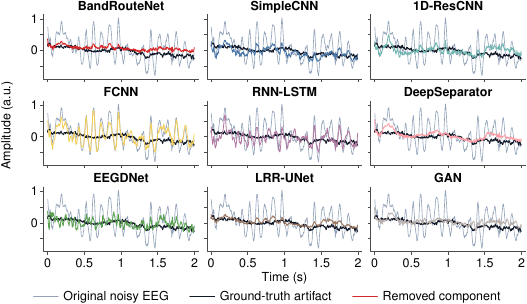}
        \end{minipage}%
    }

    \caption{Qualitative comparison on EEG-Motion
    at two evaluation levels:
    (a) Level 1 assesses preservation of the clean EEG signal,
    while (b) Level 2 assesses whether the removed component
    corresponds to the ground-truth artifact.}
    \label{fig:app_motion_qualitative}
\end{figure*}

Figures~\ref{fig:app_l1_eog}--\ref{fig:app_l2_motion} provide a qualitative view of how the denoisers transform the same contaminated input. A common vertical scale is used within each dataset so that differences in waveform shape and amplitude remain visible.

At Level~1, the overlays show whether the reconstructed waveform follows the clean EEG while correcting portions of the signal affected by contamination.
Differences among the models appear as residual artifact, excessive smoothing, amplitude attenuation, or distortion of clean temporal structure. 

Level~2 presents the complementary removal view. Comparing the estimated removed component with the true artifact reveals whether a model suppresses the appropriate temporal events with suitable morphology and magnitude.
 
Close agreement at Level~2, in turn, indicates that the output improvement is produced through the intended removal mechanism. These waveform examples therefore illustrate why SV and RF should be inspected together. They are intended to support qualitative interpretation; quantitative model comparison is provided by the complete-set \textit{TriX-EEG} metrics.


\subsection{TAAS Across Explanation Methods}

\begin{figure}[t]
    \centering
    \includegraphics{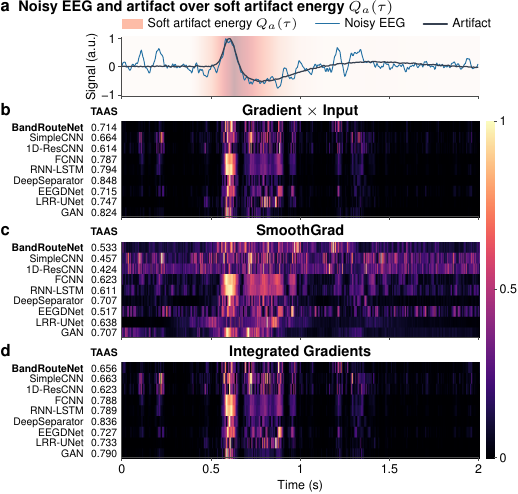}
    \caption{Temporal attribution maps on EEGDenoiseNet-EOG for nine denoisers and three explanation methods. The upper panel overlays the contaminated EEG and true artifact on the soft artifact-energy target $Q_a(\tau)$. The remaining panels show row-normalized saliency; labels report TAAS.}
    \label{fig:app_taas_methods_eog}
    \vspace{-0.3cm}
    
\end{figure}

\begin{figure}[t]
    \centering
    \includegraphics{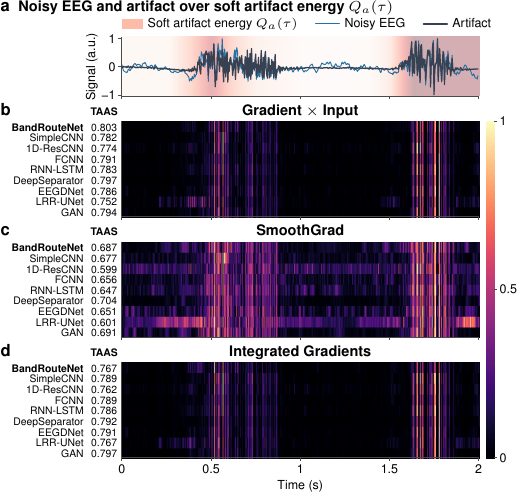}
    \caption{Temporal attribution maps on EEGDenoiseNet-EMG for nine denoisers and three explanation methods.
    }
    \label{fig:app_taas_methods_emg}
    \vspace{-0.3cm}
\end{figure}

\begin{figure}[t]
    \centering
        \includegraphics{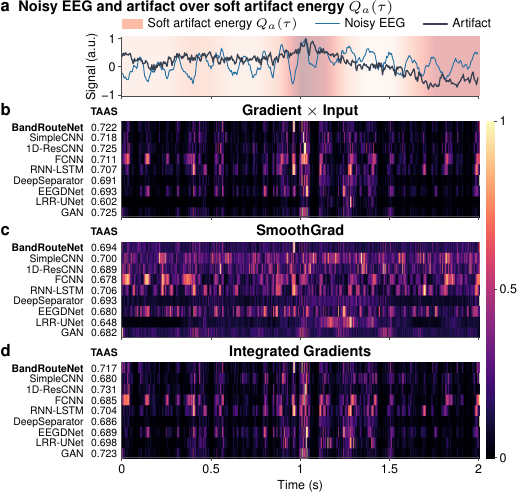}
    \caption{Temporal attribution maps on EEG-Motion for nine denoisers and three explanation methods. 
    }
    \label{fig:app_taas_methods_motion}
    \vspace{-0.3cm}
\end{figure}

The method comparison is not used to select a preferred explanation post hoc. Instead, it tests whether temporal alignment is tied to one gradient construction. Across displayed samples (Fig.~\ref{fig:app_taas_methods_eog} --~\ref{fig:app_taas_methods_motion}, Gradient $\times$ Input achieves the highest model-averaged TAAS for EEGDenoiseNet-EOG ($0.7452$; across-method spread $0.1656$) and EEGDenoiseNet-EMG ($0.7847$; spread $0.1277$), whereas Integrated Gradients performs marginally better on EEG-Motion ($0.7014$; spread $0.0159$). The small EMG and motion spreads indicate that their dataset-level conclusions are not driven by a single explanation method, whereas the larger EOG/EMG spread exposes meaningful method dependence. The model rows nevertheless differ within every panel, showing why the attribution rule must be fixed before comparison. Finally, TAAS addresses localization only: a temporally aligned map may still be non-causal or unstable, which motivates its combination with CAF and ASS in RS.

\section{Detailed Sanity-Check Results}
\label{app:sanity_checks}
We report only metrics that consume the quantity manipulated by each check. Levels 1 and 2 are deterministic for fixed test inputs and checkpoints. Stochastic Level-3 quantities are means $\pm$ sample standard deviations over five paired stratified 64-window subsets (seeds 42--46). EEGDNet~\cite{EEGDnet} is used as the representative model for these experiments.

\subsection{Random Attribution}
\begin{table}[t]
    \centering
    \caption{Random-attribution sanity check. Only attribution-dependent Level-3 metrics are shown.}
    \label{tab:app_random_attribution}
    \setlength{\tabcolsep}{3.5pt}
    \renewcommand{\arraystretch}{1.12}
    \scalebox{.65}{%
    \begin{tabular}{lccccc}
        \toprule
        \textbf{Dataset} & \textbf{Attribution} & \textbf{TAAS} & \textbf{CAF} & \textbf{ASS} & \textbf{RS} \\
        \midrule
        EEGDenoiseNet-EOG & Real & $0.5437 \pm 0.0128$ & $0.4560 \pm 0.0120$ & $0.9864 \pm 0.0004$ & $0.5700 \pm 0.0030$ \\
        EEGDenoiseNet-EOG & Shuffled & $0.4387 \pm 0.0143$ & $0.0300 \pm 0.0058$ & $0.5593 \pm 0.0023$ & $0.3309 \pm 0.0038$ \\
        \midrule
        EEGDenoiseNet-EMG & Real & $0.8188 \pm 0.0064$ & $0.4478 \pm 0.0035$ & $0.9905 \pm 0.0002$ & $0.6912 \pm 0.0028$ \\
        EEGDenoiseNet-EMG & Shuffled & $0.7769 \pm 0.0094$ & $0.0223 \pm 0.0034$ & $0.5579 \pm 0.0017$ & $0.4659 \pm 0.0033$ \\
        \midrule
        EEG-Motion & Real & $0.7497 \pm 0.0107$ & $0.1640 \pm 0.0198$ & $0.9894 \pm 0.0003$ & $0.6183 \pm 0.0085$ \\
        EEG-Motion & Shuffled & $0.7341 \pm 0.0156$ & $0.0241 \pm 0.0048$ & $0.5864 \pm 0.0048$ & $0.4792 \pm 0.0032$ \\
        \bottomrule
    \end{tabular}%
    }
    \vspace{-0.3cm}
\end{table}

\begin{table}[t]
    \centering
    \caption{Level and overall scores for trained and untrained model}
    \label{tab:app_random_model}
    \setlength{\tabcolsep}{3.5pt}
    \renewcommand{\arraystretch}{1.12}
    \scalebox{.78}{%
    \begin{tabular}{lccccc}
        \toprule
        \textbf{Dataset} & \textbf{Model} & \textbf{DS} & \textbf{FS} & \textbf{RS} & \textbf{TDS} \\
        \midrule
        EEGDenoiseNet-EOG & Trained & $0.6664$ & $0.8667$ & $0.5700 \pm 0.0030$ & $0.6905\pm 0.0012$ \\
        EEGDenoiseNet-EOG & Untrained & $0.0274$ & $0.4886$ & $0.6354 \pm 0.0089$ & $0.2041 \pm 0.0010$ \\
        \midrule
        EEGDenoiseNet-EMG & Trained & $0.5259$ & $0.8088$ & $0.6912 \pm 0.0028$ & $0.6649 \pm 0.0009$ \\
        EEGDenoiseNet-EMG & Untrained & $0.0199$ & $0.4798$ & $0.7303 \pm 0.0026$ & $0.1910 \pm 0.0002$ \\
        \midrule
        EEG-Motion & Trained & $0.3754$ & $0.4811$ & $0.6183 \pm 0.0085$ & $0.4816 \pm 0.0022$ \\
        EEG-Motion & Untrained & $0.0094$ & $0.2561$ & $0.6385 \pm 0.0056$ & $0.1154 \pm 0.0003$ \\
        \bottomrule
    \end{tabular}%
    }
\end{table}

As shown at Table~\ref{tab:app_random_attribution}, shuffling reduces every attribution-dependent metric on all datasets. Relative to the real maps, RS decreases by $42.0\%$, $32.6\%$, and $22.6\%$ for EOG, EMG, and motion, respectively. CAF is the strongest detector, falling by $85.3$--$95.0\%$, while ASS falls by $40.7$--$43.7\%$. TAAS decreases by only $5.1\%$ on EMG and $2.1\%$ on motion because a shuffled map can retain broad temporal coverage even after losing sample-wise structure. These results support a multi-criterion metrics on Level~3: TAAS tests distributional localization, CAF tests whether salient samples affect the denoising operation, and ASS tests local consistency. 


\subsection{Random or Untrained Model}

\begin{table*}[t]
    \centering
    \caption{Normalization-threshold sensitivity. Slash-separated entries correspond to the low, default, and high settings.}
    \label{tab:app_thresholds}
    \setlength{\tabcolsep}{3.5pt}
    \renewcommand{\arraystretch}{1.12}
    \scalebox{.88}{%
    \begin{tabular}{lcccccc}
        \toprule
        \textbf{Dataset} & \textbf{Threshold} & \textbf{Low/default/high} & \textbf{Utility} & \textbf{Utility values} & \textbf{Score} & \textbf{Score values} \\
        \midrule
        EEGDenoiseNet-EOG & $\Delta\mathrm{SNR}_{\max}$ & $10/20/30$ & $U_{\Delta\mathrm{SNR}}$ & $0.8973/0.6216/0.4169$ & DS & $0.7215/0.6664/0.6255$ \\
        EEGDenoiseNet-EOG & $E^{\rm BP}_{\max}$ & $0.5/1/2$ & BPS & $0.3608/0.6793/0.8396$ & DS & $0.6027/0.6664/0.6985$ \\
        EEGDenoiseNet-EOG & $E^{\rm RAER}_{\max}$ & $0.5/1/2$ & $U_{\rm RAER}$ & $0.9018/0.9509/0.9754$ & FS & $0.8569/0.8668/0.8717$ \\
        EEGDenoiseNet-EOG & $E^{\rm ARE}_{\max}$ & $0.5/1/2$ & $U_{\rm ARE}$ & $0.4761/0.7212/0.8606$ & FS & $0.8177/0.8668/0.8946$ \\
        EEGDenoiseNet-EOG & $E^{\rm AGE}_{\max}$ & $0.5/1/2$ & $U_{\rm AGE}$ & $0.8752/0.9376/0.9688$ & FS & $0.8543/0.8668/0.8730$ \\
        EEGDenoiseNet-EOG & $E^{\rm CNL}_{\max}$ & $0.5/1/2$ & $U_{\rm CNL}$ & $0.5539/0.7764/0.8882$ & FS & $0.8223/0.8668/0.8891$ \\
        \midrule
        EEGDenoiseNet-EMG & $\Delta\mathrm{SNR}_{\max}$ & $10/20/30$ & $U_{\Delta\mathrm{SNR}}$ & $0.8180/0.4782/0.3188$ & DS & $0.5938/0.5259/0.4940$ \\
        EEGDenoiseNet-EMG & $E^{\rm BP}_{\max}$ & $0.5/1/2$ & BPS & $0.2200/0.5506/0.7751$ & DS & $0.4597/0.5259/0.5708$ \\
        EEGDenoiseNet-EMG & $E^{\rm RAER}_{\max}$ & $0.5/1/2$ & $U_{\rm RAER}$ & $0.8844/0.9422/0.9711$ & FS & $0.7972/0.8088/0.8146$ \\
        EEGDenoiseNet-EMG & $E^{\rm ARE}_{\max}$ & $0.5/1/2$ & $U_{\rm ARE}$ & $0.3178/0.6349/0.8174$ & FS & $0.7454/0.8088/0.8453$ \\
        EEGDenoiseNet-EMG & $E^{\rm AGE}_{\max}$ & $0.5/1/2$ & $U_{\rm AGE}$ & $0.7471/0.8736/0.9368$ & FS & $0.7835/0.8088/0.8214$ \\
        EEGDenoiseNet-EMG & $E^{\rm CNL}_{\max}$ & $0.5/1/2$ & $U_{\rm CNL}$ & $0.3910/0.6748/0.8374$ & FS & $0.7520/0.8088/0.8413$ \\
        \midrule
        EEG-Motion & $\Delta\mathrm{SNR}_{\max}$ & $10/20/30$ & $U_{\Delta\mathrm{SNR}}$ & $0.2292/0.1664/0.1189$ & DS & $0.3879/0.3754/0.3659$ \\
        EEG-Motion & $E^{\rm BP}_{\max}$ & $0.5/1/2$ & BPS & $0.0134/0.3679/0.6839$ & DS & $0.3045/0.3754/0.4386$ \\
        EEG-Motion & $E^{\rm RAER}_{\max}$ & $0.5/1/2$ & $U_{\rm RAER}$ & $0.4425/0.6547/0.8202$ & FS & $0.4386/0.4811/0.5142$ \\
        EEG-Motion & $E^{\rm ARE}_{\max}$ & $0.5/1/2$ & $U_{\rm ARE}$ & $0.0961/0.2078/0.5053$ & FS & $0.4588/0.4811/0.5406$ \\
        EEG-Motion & $E^{\rm AGE}_{\max}$ & $0.5/1/2$ & $U_{\rm AGE}$ & $0.2421/0.5759/0.7879$ & FS & $0.4143/0.4811/0.5235$ \\
        EEG-Motion & $E^{\rm CNL}_{\max}$ & $0.5/1/2$ & $U_{\rm CNL}$ & $0.1810/0.5251/0.7614$ & FS & $0.4123/0.4811/0.5283$ \\
        \bottomrule
    \end{tabular}%
    }
    \vspace{-0.3cm}
    
\end{table*}
\begin{table}[t]
    \centering
    \caption{Wrong-artifact pairing check.}
    \label{tab:app_wrong_pairing}
    \scriptsize
    \setlength{\tabcolsep}{3.5pt}
    \renewcommand{\arraystretch}{1.12}
    \scalebox{.78}{%
    \begin{tabular}{llcccccc}
        \toprule
        \textbf{Dataset} & \textbf{Reference} & \textbf{RAS} & $U_{\rm RAER}$ & $U_{\rm ARE}$ & $U_{\rm AGE}$ & $U_{\rm CNL}$ & \textbf{FS} \\
        \midrule
        EEGDenoiseNet-EOG & True & $0.9477$ & $0.9509$ & $0.7212$ & $0.9376$ & $0.7764$ & $0.8668$ \\
        EEGDenoiseNet-EOG & Wrong raw & $0.0408$ & $0.7683$ & $0.0000$ & $0.0436$ & $0.6963$ & $0.3098$ \\
        EEGDenoiseNet-EOG & Wrong RMS-matched & $0.0408$ & $0.9509$ & $0.0000$ & $0.0401$ & $0.6963$ & $0.3456$ \\
        EEGDenoiseNet-EMG & True & $0.9185$ & $0.9422$ & $0.6349$ & $0.8736$ & $0.6748$ & $0.8088$ \\
        EEGDenoiseNet-EMG & Wrong raw & $0.0391$ & $0.7628$ & $0.0000$ & $0.0362$ & $0.6500$ & $0.2976$ \\
        EEGDenoiseNet-EMG & Wrong RMS-matched & $0.0391$ & $0.9422$ & $0.0000$ & $0.0347$ & $0.6500$ & $0.3332$ \\
        EEG-Motion & True & $0.4419$ & $0.6547$ & $0.2078$ & $0.5759$ & $0.5251$ & $0.4811$ \\
        EEG-Motion & Wrong raw & $0.0808$ & $0.1620$ & $0.0043$ & $0.0338$ & $0.5531$ & $0.1668$ \\
        EEG-Motion & Wrong RMS-matched & $0.0808$ & $0.6547$ & $0.0025$ & $0.0532$ & $0.5531$ & $0.2689$ \\
        \bottomrule
    \end{tabular}%
    }
\end{table}

\begin{table*}[t]
    \centering
    \caption{Sensitivity of Level-3 results to the attribution method.}
    \label{tab:app_xai_methods}
    \setlength{\tabcolsep}{3.5pt}
    \renewcommand{\arraystretch}{1.12}
    \scalebox{.92}{%
    \begin{tabular}{lccccc}
        \toprule
        \textbf{Dataset} & \textbf{Method} & \textbf{TAAS} & \textbf{CAF} & \textbf{ASS} & \textbf{RS} \\
        \midrule
        EEGDenoiseNet-EOG & Gradient $\times$ Input & $0.7480 \pm 0.0059$ & $0.5328 \pm 0.0124$ & $0.9869 \pm 0.0004$ & $0.6408 \pm 0.0040$ \\
        EEGDenoiseNet-EOG & Integrated Gradients & $0.7593 \pm 0.0054$ & $0.5436 \pm 0.0120$ & $0.9917 \pm 0.0001$ & $0.6475 \pm 0.0041$ \\
        EEGDenoiseNet-EOG & SmoothGrad & $0.5437 \pm 0.0128$ & $0.4560\pm 0.0120$ & $0.9864  \pm 0.0004$ & $0.5700 \pm 0.0027$ \\
        EEGDenoiseNet-EMG & Gradient $\times$ Input & $0.8571 \pm 0.0089$ & $0.4723 \pm 0.0055$ & $0.9906 \pm 0.0002$ & $0.7066 \pm 0.0032$ \\
        EEGDenoiseNet-EMG & Integrated Gradients & $0.8594 \pm 0.0078$ & $0.4787 \pm 0.0047$ & $0.9925 \pm 0.0001$ & $0.7092 \pm 0.0029$ \\
        EEGDenoiseNet-EMG & SmoothGrad & $0.8183 \pm 0.0064$ & $0.4482 \pm 0.0040$ & $0.9905 \pm 0.0000$ & $0.6912 \pm 0.0028$ \\
        EEG-Motion & Gradient $\times$ Input & $0.7472 \pm 0.0140$ & $0.1689 \pm 0.0169$ & $0.9878 \pm 0.0004$ & $0.6190 \pm 0.0093$ \\
        EEG-Motion & Integrated Gradients & $0.7471 \pm 0.0124$ & $0.1729 \pm 0.0213$ & $0.9883 \pm 0.0003$ & $0.6201 \pm 0.0101$ \\
        EEG-Motion & SmoothGrad & $0.7493 \pm 0.0106$ & $0.1635 \pm 0.0191$ & $0.9894 \pm 0.0003$ & $0.6183 \pm 0.0085$ \\
        \bottomrule
    \end{tabular}%
    }
    \vspace{-0.3cm}
\end{table*}

\begin{table}[t]
    \centering
    \caption{Clean-input identity check. 
    }
    \label{tab:app_clean_input}
    \setlength{\tabcolsep}{3.5pt}
    \renewcommand{\arraystretch}{1.12}
    \scalebox{.68}{%
    \begin{tabular}{lcccccccc}
        \toprule
        \textbf{Dataset} & \textbf{RRMSE-T} & \textbf{RRMSE-S} & \textbf{CC} &\textbf{ BPS} & \textbf{DS} & \textbf{CNL} & $U_{\rm CNL}$ & $\|\hat{a}\|_2/\|x\|_2$ \\
        \midrule
        EEGDenoiseNet-EOG & $0.3222$ & $0.3164$ & $0.9491$ & $0.7045$ & $0.6030$ & $0.1604$ & $0.8396$ & $0.3222$ \\
        EEGDenoiseNet-EMG & $0.3963$ & $0.3934$ & $0.9163$ & $0.5958$ & $0.5445$ & $0.1852$ & $0.8148$ & $0.3963$ \\
        EEG-Motion & $0.4873$ & $0.6559$ & $0.9072$ & $0.3706$ & $0.4269$ & $0.4172$ & $0.5828$ & $0.4873$ \\
        \bottomrule
    \end{tabular}%
    }
    \vspace{-0.4cm}
    
\end{table}
Table~\ref{tab:app_random_model} compares the level-wise scores of the trained
and untrained models across the three datasets. Random initialization causes
large across DS, FS, and TDS, clearly separates failure of trained and untrained models.
RS slightly
increases by $0.0203$--$0.0654$, showing that stable gradients do not certify a
useful denoising operation. This is not necessarily contradictory because RS evaluates the consistency and behavioral faithfulness of an explanation, not whether the underlying denoising behavior is desirable. These results support the conditional interpretation of
\textit{TriX-EEG}: DS and FS must establish a behavior worth explaining before
RS is considered.
\subsection{Explanation-Method Sensitivity}
\label{app:sec_explaination_methods}

We compare the effect of different explanation methods such as Integrated Gradients~\cite{gradient_integrated}, SmoothGrad~\cite{smoothgrad}, Gradient $\times$ Input on the value of proposed metrics at level 3, which are reported at Table~\ref{tab:app_xai_methods}. The RS range across methods is $0.0775$ for EOG, $0.0180$ for EMG, and $0.0018$ for motion. EOG is therefore moderately method-sensitive, principally through TAAS, whereas the EMG and motion estimates are stable at the level of uncertainty. Integrated Gradients~\cite{gradient_integrated} and Gradient $\times$ Input are close on all datasets, while SmoothGrad is lower on EOG. Consequently, explanation methods should not be treated as numerically interchangeable and should be held fixed within an evaluation.
\subsection{Clean Input with No Artifact}

As Table~\ref{tab:app_clean_input} shows, although CC remains high ($0.9072$--$0.9491$), the evaluated denoiser removes $32.22$--$48.73\%$ of the clean-signal norm and obtains only moderate DS. Correlation gives an overly favorable view when waveform amplitude or spectral content changes. $\mathrm{RRMSE}$ and BPS expose this failure most clearly on EEG-Motion, where CC is $0.9072$ but BPS is only $0.3706$. CNL provides a complementary warning: its utility is lowest on EEG-Motion ($0.5828$), indicating the strongest clean-signal distortion in the removed component after jointly controlling for the artifact. 

\subsection{Metric-Threshold Sensitivity}

As represented as Table~\ref{tab:app_thresholds}, all utilities change monotonically in the expected direction as their tolerance is relaxed, confirming that the mappings implement the intended preference. The level scores are nevertheless not threshold-invariant: the largest DS spread is $0.1341$ and occurs for BPS, whereas the largest FS spread is $0.1160$ and occurs for $U_{\rm CNL}$ on EEG-Motion. $U_{\rm AGE}$ and $U_{\rm ARE}$ are also comparatively sensitive on this dataset, while $U_{\rm RAER}$ produces smaller changes. These results distinguish robustness of direction from invariance of magnitude: the metrics respond correctly, but their calibration affects the numerical score. Thus, thresholds must therefore be fixed before model comparison and justified using domain tolerances.

\subsection{Wrong Artifact Pairing}

We replace the true artifact with a randomly paired artifact from another type and also evaluate RMS-matched wrong artifacts. As shown in Table~\ref{tab:app_wrong_pairing}, RMS matching leaves RAER unchanged by construction, but RAS collapses, $U_{\rm ARE}$ falls toward $0$, and $U_{\rm AGE}$ falls to $0.0401$, $0.0347$, and $0.0532$ on EOG, EMG, and EEG-Motion, respectively. Accordingly, FS decreases by $0.5212$, $0.4756$, and $0.2122$, providing a substantially clearer rejection of the wrong artifact. CNL is not required to change monotonically under this intervention---on EEG-Motion its utility increases slightly from $0.5251$ to $0.5531$---because changing the artifact regressor can redistribute variance between correlated predictors.
The check should therefore be interpreted from the joint metric pattern and FS.

\subsection{Summary of Sanity-Check Evidence}

Overall, the sanity checks highlight the complementary diagnostic roles and limitations of the three evaluation levels. Level~1 identifies ineffective denoisers and distortions introduced to clean inputs; Level~2 characterizes whether the removed component corresponds to the reference artifact and indicates clean-signal-aligned leakage; and Level~3 examines distinct properties of explanation behavior, including artifact alignment, intervention-based relevance, and stability. The controlled failure cases further demonstrate that individual metrics cannot establish denoising quality in isolation: high waveform correlation does not guarantee neural preservation, appropriate removed-signal energy does not establish artifact correspondence, and stable attributions do not imply effective artifact removal. These findings motivate the layered evaluation design of \textit{TriX-EEG}.

Although TDS provides a convenient summary ranking, it should be interpreted alongside the level-wise scores and component metrics to identify specific failure modes. In particular, ER should be assessed in conjunction with SV and RF, while artifact-identity measures remain essential for interpreting removed-signal characteristics. Normalization thresholds and attribution configurations should also be specified consistently to support reproducible comparisons.